\documentclass[amsmath, amssymb, aps, prx, twocolumn, superscriptaddress, footinbib]{revtex4-2}
\usepackage{mathtools}
\usepackage{mathrsfs, amsmath, wasysym}
\usepackage{bbold}
\usepackage[mathscr]{eucal}
\usepackage{xfrac}
\usepackage{graphicx}
\usepackage{dcolumn}
\usepackage{bm}
\usepackage{color}
\usepackage{tabularx}
\usepackage{comment}
\usepackage{natbib}
\usepackage[colorlinks,linkcolor=blue,citecolor=blue,urlcolor=blue]{hyperref}
\usepackage[normalem]{ulem}
\usepackage[all]{hypcap}
\usepackage[dvipsnames,usenames,table]{xcolor}
\usepackage{braket,mleftright}

\newcommand{\nocontentsline}[3]{}
\newcommand{\tocless}[2]{\bgroup\let\addcontentsline=\nocontentsline#1{#2}\egroup}

\newcommand{\bk}{{\bf k}}

\newcommand{\bq}{{\bf q}}

\newcommand{\bz}{{\bf z}}

\newcommand{\ba}{{\bf a}}

\newcommand{\bD}{{\bf D}}

\newcommand{\bS}{{\bf S}}

\newcommand{\jAF}{J_{\text{AF}} }

\newcommand{\bsigma}{\boldsymbol{\sigma} }

\newcommand{\bdelta}{\boldsymbol{\delta} }
\newcommand{\blambda}{\boldsymbol{\lambda} }

\newcommand{\be}{\begin{equation}}
\newcommand{\ee}{\end{equation}}
\newcommand{\beg}{\begin{gather}}
\newcommand{\eeg}{\end{gather}}
\newcommand{\beq}{\begin{eqnarray}}
\newcommand{\eeq}{\end{eqnarray}}
\newcommand{\bea}{\begin{align}}
\newcommand{\eea}{\end{align}}
\newcommand{\beqq}{\begin{eqnarray*}}
\newcommand{\eeqq}{\end{eqnarray*}}

\newcommand{\up}{\uparrow}
\newcommand{\down}{\downarrow}

\begin{document}

\title{Canonical strong coupling spin wave expansion of Kondo lattice magnets. \\
II. Itinerant ferromagnets and topological magnon bands}

\author{M. Frakulla}
\affiliation{Department of Physics, Drexel University, Philadelphia, PA 19104, USA}%

\author{J. Strockoz}
\affiliation{Department of Physics, Drexel University, Philadelphia, PA 19104, USA}%

\author{D. Antonenko}
\affiliation{Department of Physics, Drexel University, Philadelphia, PA 19104, USA}%
\affiliation{Department of Physics, Yale University, New Haven, CT 06520, USA}%

\author{J. W. F. Venderbos}
\affiliation{Department of Physics, Drexel University, Philadelphia, PA 19104, USA}%
\affiliation{Department of Materials Science \& Engineering, Drexel University, Philadelphia, PA 19104, USA}%

\begin{abstract}
In this paper we apply the canonical spin wave theory developed for itinerant Kondo lattice magnets in the strong coupling regime to Kondo ferromagnets, and address two general questions pertaining to their magnetic excitations. First, we compute corrections to the strong coupling (i.e., double-exchange) spin wave dispersion of itinerant ferromagnets. We show that the spin wave dispersion beyond the strong coupling limit can be mapped to the spin wave dispersion of a Heisenberg ferromagnet with farther neighbor exchange couplings, and discuss how this affects instabilities towards antiferromagnetism. Second, we examine the effect of including electronic spin-orbit coupling in the spin wave theory of Kondo ferromagnets. Including spin-orbit coupling is natural and straightforward in the formulation of the canonical spin wave expansion. Our key result is to demonstrate that the linear spin wave Hamiltonian of the itinerant Kondo ferromagnet can be mapped to the spin wave Hamiltonian of a Heisenberg ferromagnet with easy-axis Ising anisotropy and antisymmetric Dzyaloshinskii-Moriya exchange interaction. We show that in the case of the Kane-Mele honeycomb lattice Kondo ferromagnet this leads to topological magnon bands, and discuss the implications of this result for itinerant ferromagnets more broadly.
\end{abstract}

\date{\today}

\maketitle

\section{Introduction \label{sec:intro}}

The study of low-energy excitations in ordered magnets provides key information on the nature of the magnetic ground state, as well as the exchange interactions which give rise to such ground states. A well-established and broadly deployed experimental technique for measuring such spin wave excitations is inelastic neutron scattering, which has become an indispensable tool in the study of magnetic materials~\cite{Lovesey1984,Shirane2002}.  

A number of different theoretical approaches are available to model and interpret the spin wave spectra obtained from experiment. The most commonly used class of models for studying spin systems are Heisenberg models, which describe include pairwise interactions between spins on a lattice. Such models may include next-nearest or farther-neighbor interactions, magnetocrystalline anisotropies, and higher-order interaction terms (e.g. biquadratic interactions). Within such Heisenberg models, spin excitations are often analyzed in the quasiclassical limit of large spins, $S \gg 1$, which provides a good or even excellent description of the low-energy excitations of ordered magnets~\cite{Fishman2018}. The standard technical approach for such analysis is to bosonize the spins using the Holstein-Primakoff prescription~\cite{Holstein:1940p1098, Auerbach1994} and then performing a $1/S$ spin wave expansion. To lowest order this yields a Hamiltonian for non-interacting spin waves, from which the semiclassical spin wave dispersion can be determined. Higher order terms in the expansion give rise to boson interactions and can be used to determine quantum corrections to the semiclassical dispersion or magnon decay processes~\cite{ZhitomirskyRevModPhys85}.

A different class of magnetic systems instead requires a description which includes both local moment spins and itinerant charge carriers. The description of spin wave excitations in such itinerant magnets is generally more challenging due to the interactions between the spins and the itinerant electrons. In the preceding companion paper we have developed a strong coupling spin wave expansion for the class of itinerant Kondo lattice magnets~\cite{strockoz2024}, which are described by a Kondo lattice Hamiltonian of the general form
\be  \label{eq:H_klm}
H = \sum_{ij} t_{ij} c^\dagger_i c_j -  \frac{J_K}{2S}\sum_i \bS_i \cdot c^\dagger_i \bsigma c_i + H_J
\ee
Here $ c_{i\sigma}$ are the electron operators with $\sigma=\up,\down$ and $\bS_i$ are the local spin-$S$ moments. The first term in \eqref{eq:H_klm} describes the hopping of the itinerant electrons and the second term is the on-site Kondo coupling with coupling constant $J_K$. The third term $H_J$ accounts for a direct Heisenberg exchange interaction between the local moments, as opposed to the indirect interaction generated via the mobile electrons, and takes the form $H_J = (1/2S^2) \sum_{ij}J_{ij} \bS_i\cdot \bS_j $, with exchange couplings $J_{ij}$ between pairs of spins. 

The spin wave expansion developed in the companion paper starts from the assumption that the Kondo coupling $J_K$ is the largest energy scale, and is therefore a spin wave expansion constructed for the strong coupling regime. In the classical limit of Eq.~\eqref{eq:H_klm}, i.e., when the spins are treated as classical vectors, the strong coupling assumption implies a large energetic separation between electrons whose spin is aligned and anti-aligned with the classical local moments and motivates constructing an effective theory for the low-energy degrees of freedom only. The spin wave expansion does exactly that, by first performing the Holstein-Primakoff substitution for the spins (assuming an ordered classical ground state) and then using a canonical transformation to remove all terms from the Hamiltonian which connect the high- and low-energy subspaces, i.e., remove all electron spin-flip terms. The use of a canonical unitary transformation is a powerful method for obtaining strong coupling expansions of strongly correlated systems~\cite{Schrieffer:1966p491,MacDonald:1988p9753,Bravyi:2011p2793,Chernyshev:2004p235111}, and in this regime it yields a systematic $1/S$ expansion of the Kondo lattice Hamiltonian. In addition, it also produces a systematic perturbative expansion in $t_{ij}/J_K$ and $J_{ij}/J_K$. A remarkable feature of the spin wave expansion is that, after projecting out the high-energy states, the resulting effective Hamiltonian describes spinless fermions in a state of total spin $S\pm 1/2$ (depending on the sign of $J_K$) and bosons which correspond to fluctuations of total spin. These are the natural quantum degrees of freedom in the strong coupling regime, i.e., when $J_K$ is very large. 

The purpose of this paper is to examine two applications of the canonical spin wave expansion. First, we consider itinerant ferromagnets in the strong coupling limit ($J_K\rightarrow \infty$) and determine the corrections to spin wave dispersion to first order in the strong coupling parameter $t/J_K$. As a second application, we then study the effect of electronic spin-orbit coupling on the spin wave dispersion of itinerant ferromagnets. 

Previous work addressing the spin wave dispersion of itinerant Kondo lattice magnets has primarily focused on ferromagnets~\cite{Kubo:1972p21,Furukawa:1996p1174,Nagaev:1998p827,Golosev:2000p3974,Perkins:1999p1182,Shannon:2002p104418, MoreoRepProgPhys2025, Wang:1998p7427}, motivated by the experimental observation of itinerant ferromagnetism~\cite{Dagotto:2001p1,Izyumov:2001p109}. In the strong coupling limit, sometimes also referred to as the double-exchange regime~\cite{Zener:1951p440,Anderson:1955p675, deGennesPhysRev118,Dagotto:2001p1,Izyumov:2001p109,Vogt_2001}, the spin wave dispersion of an itinerant Kondo ferromagnet was shown to be formally equal to the spin wave dispersion of a nearest-neighbor Heisenberg ferromagnet~\cite{ Kubo:1972p21, Furukawa1994JPSJ, Perkins:1999p1182, Golosev:2000p3974, Shannon:2002p104418, Shannon:2002p235}, with an effective ferromagnetic exchange coupling given by the average kinetic energy of the electrons. Here we leverage the canonical spin wave expansion to go beyond the strong coupling limit and show that, when corrections to the strong coupling limit are taken into account, the spin wave dispersion of itinerant ferromagnets can be mapped to a Heisenberg model with farther-neighbor exchange couplings (e.g., a $J_1$-$J_2$-$J_3$ Heisenberg model). Importantly, these corrected effective exchange couplings may be ferromagnetic or antiferromagnetic depending on electron density, and may be compatible or frustrated. We demonstrate in particular that the effective nearest neighbor exchange coupling changes from ferromagnetic to antiferromagnetic as the electron density increases and approaches commensurate fillings. 

The upshot of the mapping to an effective Heisenberg model with multiple exchange couplings is that it removes a spurious collapse of the spin wave dispersion (in the strong coupling double-exchange limit) when an explicit antiferromagnetic exchange coupling $\jAF$ between the spins [as given by $H_J$ in Eq.~\eqref{eq:H_klm}] reaches a critical strength. Such a collapse suggests a large unphysical degeneracy, which we show is lifted when corrections to the strong coupling limit are included. 

We demonstrate the mapping to an effective Heisenberg model with extended pairwise interactions first in the case of a simple toy model in one dimension, and then generalize this result to the two-dimensional square and triangular lattices.  

In the second part of this paper we examine the effect of spin-orbit coupling. An appealing feature of the canonical spin wave expansion is that the effect of spin-orbit coupling can be included in a natural and straightforward way. The role of spin-orbit coupling in itinerant Kondo lattice magnets has received considerable attention as source of frustration, and therefore as a mechanism for stabilizing unconventional magnetic ground states such as Skyrmion crystals~\cite{Hayami:2018p137202,Okada:2018p224406,Banerjee:2014p031045,Meza:2014p085107,Zhang:2020p024420,Kathyat:2020p075106,Kathyat:2021p035111}. The implications of spin-orbit coupling for the spin wave dispersion of itinerant Kondo magnets have not been addressed, to the best of our knowledge. In this paper, we develop a spin wave theory for itinerant ferromagnets in the strong coupling limit in the presence of spin-orbit coupling. We specifically consider two models, a zigzag chain in one dimension and the honeycomb lattice in two dimensions, in which a spin-orbit coupling of the Kane-Mele type~\cite{Kane:2005p226801} arises for the electrons. Our central result is to demonstrate that---at the level of linear spin wave theory--- the spin wave dispersion of these itinerant spin-orbit coupled ferromagnets is formally equal to the spin wave dispersion of a Heisenberg spin model with Ising and DM exchange anisotropies. The effective Ising and DM anistropies are calculated and their dependence on electron density is examined.

In the case of the honeycomb lattice, this has the important consequence that the magnon bands are topological and have a Chern number. Topological magnon bands and topological magnon insulators have attracted attention as realizations of nontrivial topology in systems with bosonic quasiparticles~\cite{Malki:2020p20003,Kondo:2020pptaa151,McClarty:2021p1,Li:2021p1,Bonbien:2022p103002}, and as a venue for probing thermal Hall effects~\cite{Katsura:2010p066403,Matsumoto:2014p054420}. In particular, previous work on honeycomb lattice Heisenberg ferromagnets has shown that the DM interaction enters as a Haldane term \cite{Haldane:1988p2015} in the linear spin wave Hamiltonian, leading to a topological gap between the two magnon branches~\cite{Owerre:2016p386001,KimPRL117,Owerre:2016p043903,McClarty:2021p1}. Here we demonstrate that such topological magnons also arise in spin-orbit coupled itinerant Kondo ferromagnets. 

%
%
%
%
%
%
%
%

\section{Ferromagnets beyond strong coupling}
 \label{sec:FM}

As a first application of the canonical spin wave expansion, we revisit the spin wave excitations of Kondo lattice ferromagnets. As mentioned in the introduction, historically the magnetic excitations of itinerant ferromagnets have received the most attention, due in large part to the experimental realization of itinerant ferromagnetism in the colossal magnetoresistance manganites~\cite{Dagotto:2001p1,Furukawa:1996p1174,Izyumov:2001p109,TokuraRepProgPhys69}, as well as in other correlated systems. Especially in the limit of low carrier density, ferromagnetism appears to be a common feature of itinerant magnets described by the Kondo lattice model \cite{Tsunetsugu:1997p809, Tsunetsugu:1993p8345, Sigrist:1991p2211, Sigrist:1992p175, frakulla2024}. 

We begin this section with a brief discussion of the ferromagnetic spin wave spectrum in the strong coupling limit. It has been previously pointed out that, in this limit, the spin wave spectrum is formally equal to that of a Heisenberg ferromagnet, with an effective ferromagnetic exchange coupling proportional to the average kinetic energy per bond. This leads to the peculiar result that the spin wave spectrum vanishes when a competing nearest neighbor antiferromagnetic Heisenberg coupling becomes of equal strength, indicating an artificially large degeneracy at the classical level.  Here we examine the $t/J_K$ corrections to the strong coupling limit and show how these corrections lift this degeneracy. More generally, we show that including $t/J_K$ corrections gives rise to a spin wave dispersion resembling that of a Heisenberg ferromagnet with effective exchange couplings beyond nearest neighbors. This exposes how the strong coupling regime may be connected to the weak-coupling RKKY limit~\cite{Ruderman:1954p99, Kasuya:1956p45, Yosida:1957p893}. 

Our analysis in this section starts from a version of the Kondo Hamiltonian of Eq.~\eqref{eq:H_klm} which includes only nearest neighbor couplings, i.e., nearest neighbor hopping $t$ and nearest neighbor antiferromagnetic exchange coupling $\jAF$. We focus on Bravais lattices and discuss specific examples in one and two dimensions. We furthermore assume a ferromagnetic Kondo coupling, $J_K>0$, in which case the strong coupling limit is generally referred to as the double-exchange limit. 

\subsection{Ferromagnets in the strong coupling limit \label{ssec:FM-DE}}

Consider a Kondo lattice ferromagnet in the strong Kondo coupling limit on a general Bravais lattice in any dimension. In this limit, the low-energy electrons are in a state of total spin $S+1/2$ at each site, which classically corresponds to electrons whose spin is fully aligned with the local moment. To describe these effectively spinless electrons, we introduce the operators $f_i$. The Hamiltonian $\mathcal H_f$ is then given by 
\be
\mathcal H_f =  \sum_{ij}( t_{ij} -\mu \delta_{ij}) f^\dagger_{i}f_{j}, \label{eq:H_f}
\ee
which is diagonalized by Fourier transformation and yields $\mathcal H_f =  \sum_{\bk}( \varepsilon_{\bk} -\mu) f^\dagger_{\bk}f_{\bk}$. Here $\varepsilon_{\bk} = -zt \gamma_\bk$ is the fermion dispersion, $z$ is the coordination number, and $\gamma_\bk  = (1/z) \sum_{\bdelta} e^{i \bk\cdot\bdelta}$; the sum is over all nearest neighbors $\bdelta$. The chemical potential is $\mu$.

The leading order spin wave spectrum (i.e., determined in linear spin wave theory) is obtained from the effective spin wave Hamiltonian. The ferromagnetic spin wave Hamiltonian in the strong coupling limit, denoted  $\mathcal H_0$, is given by
\be
\mathcal H_0 = \frac{1}{4S} \sum_{ij} (t_{ij} f^\dagger_{i}f_{j} +2 J_{ij})(2 a^\dagger_ia_j - a^\dagger_ia_i- a^\dagger_ja_j) , \label{eq:H_0-FM}
\ee
which should be understood as a sum over two contributions: a contribution generated from the Kondo coupling and the familiar contribution from the Heisenberg coupling. Recall that here we take $t_{ij}= -t$ and $J_{ij} = \jAF$ for nearest neighbor pairs $\langle ij \rangle$ and zero otherwise. The linear spin wave dispersion of the bosons is obtained by taking the average over the fermions with respect to the ground state of \eqref{eq:H_f}. This yields
\be
\widetilde{\mathcal H}_{0} =   \frac{z(\tilde J_1 -\jAF)}{S} \sum_{\bq}(1-\gamma_\bq) a^\dagger_\bq a_\bq . \label{eq:H_a_eff}
\ee
Here $\tilde J_1 $ is an effective nearest neighbor ferromagnetic exchange coupling generated by the coupling to the electrons and is given by
\be
 \tilde J_1\equiv \frac12 c_1 t , \qquad c_1 =  \frac{1}{N} \sum_{\bk} \gamma_{\bk} n_\bk,  \label{eq:tildeJ1}
\ee
where $n_\bk=\langle f^\dagger_{\bk}f_{\bk} \rangle$. The spin wave dispersion $\omega^{(0)}_\bq$ in the strong coupling limit then readily follows, and is given by
\be
S \omega^{(0)}_\bq = z(\tilde J_1 -\jAF)(1-\gamma_\bq). \label{eq:omega_q_DE}
\ee
This reproduces a well-known result and implies that, at the level of linear spin wave theory, a Kondo lattice ferromagnet in the strong coupling limit is equivalent to a Heisenberg ferromagnet with an exchange coupling $\tilde J_1 -\jAF$, the sum of an effective ferromagnetic exchange coupling $\tilde J_1$ and the antiferromagnetic coupling $\jAF$. Note that $\tilde J_1$ can be expressed in terms of the average kinetic energy per site $\varepsilon_{\text{FM}}$ as $\tilde J_1 =-\varepsilon_{\text{FM}}/2z $ and clearly depends on the density of carriers.

An immediate consequence of Eq.~\eqref{eq:omega_q_DE} is that the ferromagnet becomes unstable as $\jAF$ reaches the value $\tilde J_1 $. In fact, the dispersion vanishes completely when $\tilde J_1=\jAF$, which suggests that at this value the competition between $\tilde J_1$ and $\jAF$ gives rise to a large degeneracy of classical ground states. This is an artifact of the strict strong coupling limit and is not expected to survive the inclusion of $t/J_K$ corrections. We now show that when these corrections are included, the spectrum does not vanish and the instability occurs only at specific wave vectors, as is usual in spin models when the classical phase boundary is reached.

\subsection{Including $t/J_K$ corrections \label{ssec:FM-correct} }

The first order $t/J_K$ corrections to the linear spin wave theory of Kondo ferromagnets are described by the Hamiltonian \cite{strockoz2024}
\begin{multline}
 \mathcal H_{1} = \frac{1}{4S J_K }  \sum_{ijk} t_{ik} t_{kj}  \left[2 a^\dagger_ia_k + 2a^\dagger_ka_j- 2   a^\dagger_i a_j  \right.  \\
\left. - (a^\dagger_ia_i +a^\dagger_j a_j ) \right]  f^\dagger_{i}f_{j}, \label{eq:tJ_K-correct}
\end{multline}
such that the full Hamiltonian becomes $\mathcal H_{0}+\mathcal H_{1}  $. Taking the average over the fermions and then Fourier transforming, we find 
\be
\widetilde{\mathcal H}_{1} =  -\frac{z^2 t^2}{2J_K SN}  \sum_{\bk,\bq} (\gamma_{\bk+\bq} - \gamma_{\bk})^2  n_{\bk}  a^\dagger_\bq a_\bq, \label{eq:H_a_eff_correct}
\ee
which applies to a general Bravais lattice in any dimension. It follows that the correction to the spin wave dispersion, denoted $ \omega^{(1)}_\bq$, is given by
\be
S \omega^{(1)}_\bq  =  -\frac{z^2 t^2}{2J_KN}  \sum_{\bk} (\gamma_{\bk+\bq} - \gamma_{\bk})^2  n_{\bk},\label{eq:omega_q_DE_correct}
\ee
which is in full agreement with the result of Furukawa~\cite{Furukawa:1996p1174}.

We now proceed to examine the corrections to the spin wave dispersion in more detail. We first consider the simplest case of a one-dimensional (1D) ferromagnet and then discuss how the key features of such a toy model generalize to the square and triangular lattices in two dimensions (2D).  

\begin{figure}
	\includegraphics[width=0.65\columnwidth]{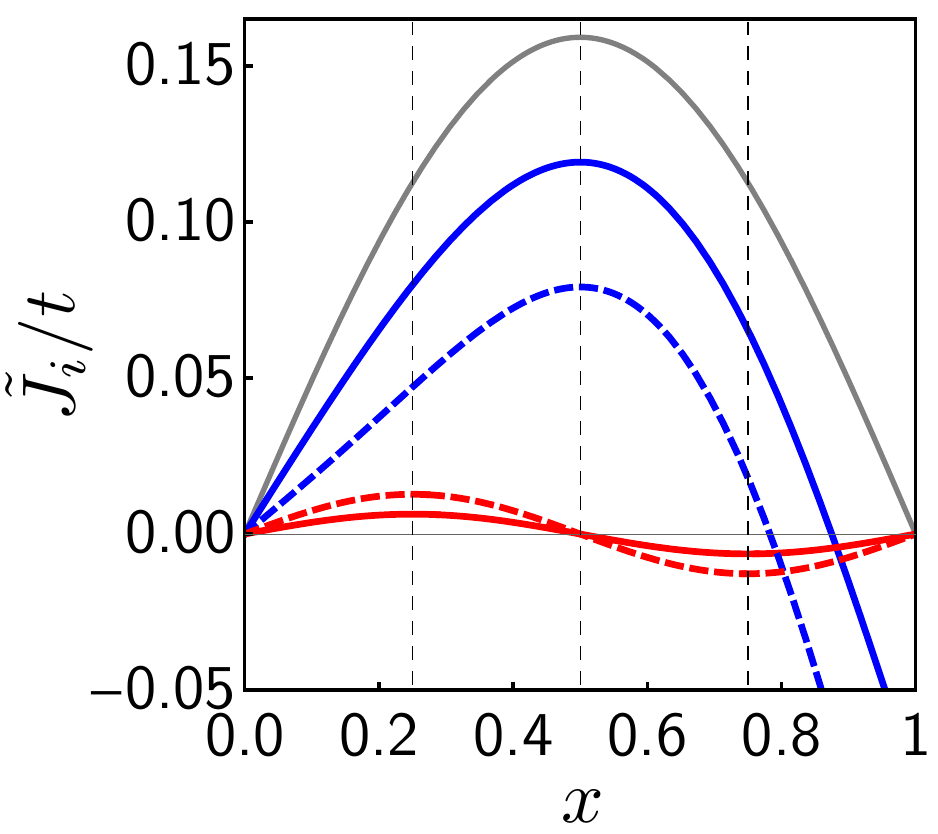}
	\caption{Effective exchange couplings $\tilde J_1$ and $\tilde J_2$ of the 1D Kondo ferromagnet, as given by Eq.~\eqref{eq:J1-J2-1D}. Solid blue and red curves correspond to $\tilde J_1$ and $\tilde J_2$ for $t/J_K=0.08$; the dashed curves correspond to $t/J_K=0.16$. For comparison, the grey solid curve corresponds to $\tilde J_1$ in the strong coupling limit ($t/J_K=0$), as given by Eq.~\eqref{eq:tildeJ1}. Here $x$ is the electron filling, i.e., the number of electrons per site.}
	\label{fig:J_i-1D}
\end{figure}

\subsection{Toy model in one dimension \label{ssec:1D} }

In the case of a simple 1D Kondo ferromagnet, the full spin wave dispersion including $t/J_K$ corrections can be shown to take the form
\begin{align}
S \omega_q  &= \omega^{(0)}_q+\omega^{(1)}_q\\
& = z(\tilde J_1 -\jAF)(1-\gamma_q) +   z \tilde J_2 (1-\gamma'_q) , \label{eq:omega_q_1D}
\end{align}
where $\gamma'_q =  \cos 2q$ is the second-nearest neighbor hopping form factor; $\gamma_q = \cos q$ is the 1D version of the form factor defined below Eq.~\eqref{eq:H_f}. Note that for a general Bravais lattice the second-nearest neighbor hopping form factor is defined as $\gamma'_\bq  = (1/z) \sum_{\bdelta'} e^{i \bq\cdot\bdelta'}$ (here $\bdelta'$ are the second-nearest neighbor vectors). 

In Eq.~\eqref{eq:omega_q_1D} the effective exchange couplings $\tilde J_1$ and $\tilde J_2$ are defined as
\be
\tilde J_1 = \frac12 c_1 t - \frac{t^2}{J_K}(c_0+c_2) , \qquad \tilde J_2=\frac{t^2}{2J_K}c_2, \label{eq:J1-J2-1D}
\ee
where $c_1$ was introduced in \eqref{eq:tildeJ1} and $c_{0,2}$ are given by
\be
c_0 = \frac{1}{N}  \sum_{k}  n_k, \qquad c_2 = \frac{1}{N}  \sum_{k} \gamma'_k\, n_k. \label{eq:c_i-1D}
\ee
The form of $\omega_q$ in Eq.~\eqref{eq:omega_q_1D} is identical to the spin wave dispersion of a Heisenberg ferromagnet with first \emph{and} second nearest neighbor exchange couplings $\tilde J_1 -\jAF$ and $\tilde J_2$, respectively. Note that, given their definition in Eq.~\eqref{eq:omega_q_1D}, these exchange couplings are ferromagnetic when they are positive (and antiferromagnetic when they are negative). Hence, at the level of leading order \emph{linear} spin wave theory, Eq.~\eqref{eq:omega_q_1D} establishes a connection between a Kondo ferromagnet and a Heisenberg ferromagnet with exchange couplings beyond first nearest neighbors. We observe that the $t/J_K$ corrections reduce $\tilde J_1$ from its value in the strong coupling limit \eqref{eq:tildeJ1}.

The effective couplings $\tilde J_1$ and $\tilde J_2$ depend on the filling fraction $x$ via the integrals $c_{0,1,2}$. Here $x$ is understood as the number of electrons per site. In 1D the integrals are readily calculated and we find $c_1 = \sin(\pi x)/\pi$ and $c_2 = \sin(2\pi x)/2\pi$; $c_0 = x$ by definition. The resulting exchange couplings $\tilde J_{1,2}$ are shown in Fig.~\ref{fig:J_i-1D}  (blue and red, respectively) for $t/J_K=0.08,0.16$ (solid and dashed curves, respectively). The gray curve corresponds to $\tilde J_{1}$ in the strong coupling limit, in which case $\tilde J_{1}/ t = c_1/2$. 

Two key features of the effective exchange couplings are apparent in Fig.~\ref{fig:J_i-1D}. The first is that the second-nearest neighbor coupling $\tilde J_{2}$ is ferromagnetic for filling fractions $x<1/2$ and is antiferromagnetic for $x>1/2$. This has important implications for the instability of the ferromagnet as the direct antiferromagnetic coupling $\jAF$ is increased. The nature of this instability can be analyzed in two equivalent ways: by determining the appearance of additional zero modes in the spin wave dispersion $\omega_q$ of Eq.~\eqref{eq:omega_q_1D}, or by minimizing the classical energy of a Heisenberg spin model with (ferromagnetic) nearest neighbor exchange coupling $\tilde J_{1}-\jAF$ and next-nearest neighbor exchange coupling $\tilde J_{2}$. Taking the former approach leads to the conclusion that for $x<1/2$, i.e., when $\tilde J_{2}$ is ferromagnetic, the spin wave energy vanishes at $q=\pi$ when $\tilde J_{1}=\jAF$, signaling an instability towards the antiferromagnet. Instead, when $x>1/2$ and $\tilde J_{2}$ is antiferromagnetic, new zero modes appear in the vicinity of $q=0$, which signals an instability towards a long-wavelength spin spiral state. This instability is marked by a vanishing of the spin stiffness of the quadratic Goldstone modes and therefore occurs at  $\tilde J_{1}-\jAF = 4 |\tilde J_{2}|$. 

This analysis shows that the $t/J_K$ corrections to the strong coupling limit remove the aforementioned artifact of a collapsing spin wave dispersion at $\tilde J_{1}=\jAF$ [see Eq.~\eqref{eq:omega_q_DE}], which in turn implies that they remove the (artificial) degeneracy implied by such collapse. Note that $x=1/2$ remains special, since $\tilde J_{2}=0$ in this case.

Another key feature is the sign change of $\tilde J_{1}$ from ferromagnetic to antiferromagnetic as $x$ increases, reaching a limiting value of $\tilde J_{1} = -t^2/J_K$ when $x = 1$ (i.e., at one electron per site). The value of effective nearest neighbor coupling at $x=1$ follows readily from the fact that $c_0=1$ and $c_1=c_2=0$ when $x=1$. This transition from ferromagnetic to antiferromagnetic coupling reflects a known property of the strong coupling Kondo model (or double-exchange model), which is the shift from favoring ferromagnetism to favoring antiferromagnetism when the band filling reaches one electron per site~\cite{Kathyat:2020p075106,Mukherjee:2021p134424,strockoz2024}. At $x=1$ all sites are occupied and electron hopping---the origin of ferromagnetism in the strong coupling limit~\cite{Zener:1951p440,Anderson:1955p675, deGennesPhysRev118,Izyumov:2001p109}---cannot occur. Only virtual processes via the high energy spin states are possible, which are perturbative in $t/J_K$ and generate an antiferromagnetic coupling of strength $t^2/J_K$.

\begin{figure}
	\includegraphics[width=\columnwidth]{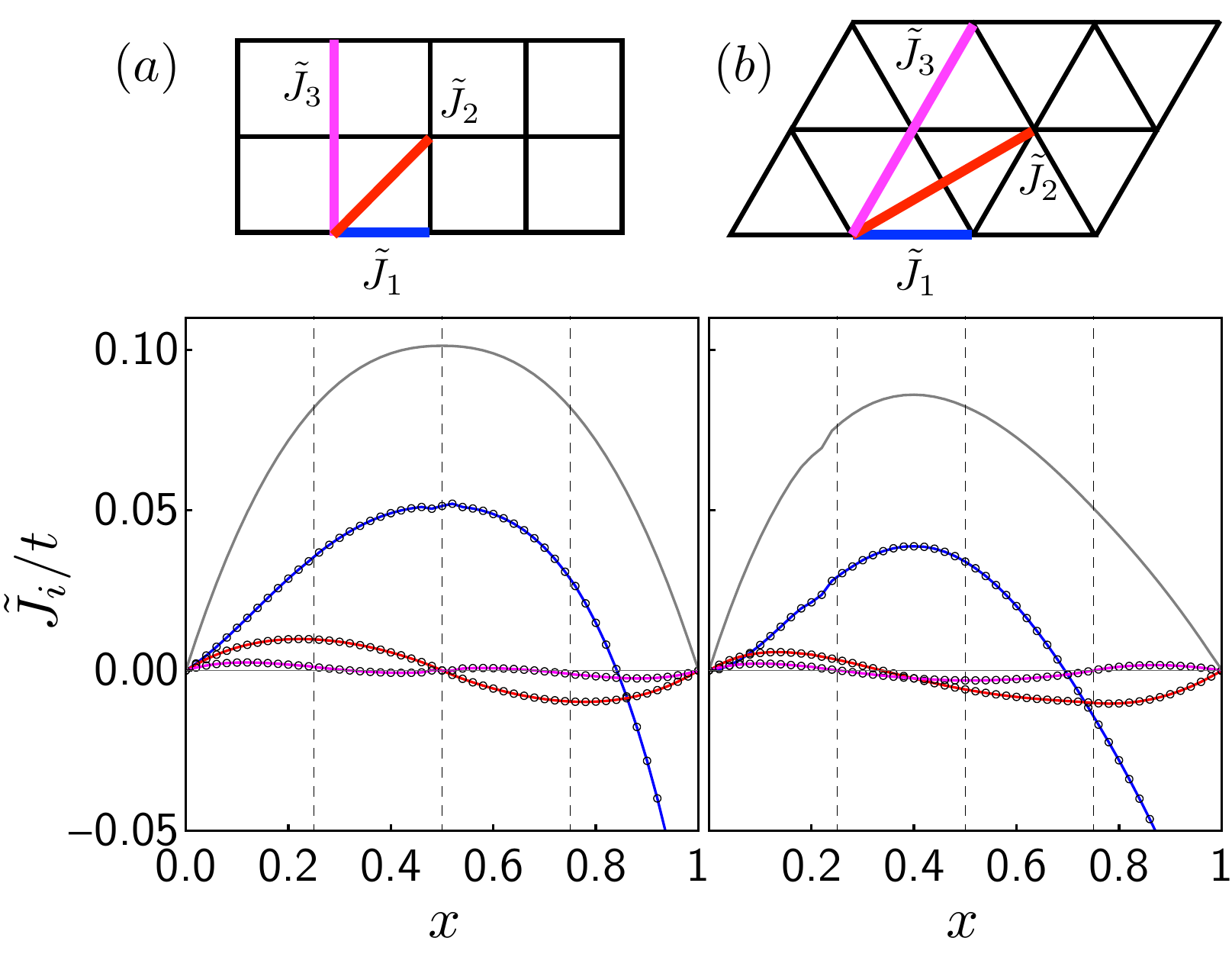}
	\caption{Effective exchange couplings $\tilde J_{1,2,3}$ as a function of filling fraction $x$. Shown are the (a) square lattice and (b) triangular lattice couplings, as given by Eqs.~\eqref{eq:J1-square},~\eqref{eq:J2-J3-square}, and~\eqref{eq:J1-tri}. Here the blue, red, and magenta curves correspond to $\tilde J_{1}$, $\tilde J_{2}$, and $\tilde J_{3}$, respectively, and we have taken $t/J_K = 0.1$. For comparison, the gray curve corresponds to the strong coupling limit of $\tilde J_1$, i.e., $c_1 t/2$.}
	\label{fig:J_i-2D}
\end{figure}

\subsection{Square and triangular lattice ferromagnets \label{ssec:2D} }

Consider next the spin wave dispersion of a square lattice and triangular lattice Kondo ferromagnet. In both cases, we find that $\omega_\bq  $, i.e., the sum of \eqref{eq:omega_q_DE} and \eqref{eq:omega_q_DE_correct}, takes the general form
\be
S \omega_\bq = z(\tilde J_1 -\jAF)(1-\gamma_\bq) +   z \tilde J_2 (1-\gamma'_\bq) +   z \tilde J_3 (1-\gamma''_\bq), \label{eq:omega_q_2D}
\ee
where $\gamma''_\bq  = (1/z) \sum_{\bdelta''} e^{i \bq\cdot\bdelta''}$, with $\bdelta''$ the third-nearest neighbor vectors.  Consistent with the toy model in 1D, this shows that $\omega_\bq$ is identical in form to the spin wave dispersion of a Heisenberg ferromagnet with effective exchange couplings $\tilde J_{1,2,3}$. For both lattices the $t/J_K$ corrections generate both a second- and a third-nearest neighbor effective exchange coupling. 

As in the 1D model, the effective exchange couplings can be expressed in terms of filling-dependent integrals $c_i$. Explicit expressions for these integrals are given in Appendix~\ref{app:integrals}. For the square lattice we find that the effective first nearest neighbor coupling takes the form
\be
 \tilde J^\Box_1=  \frac12 c_1t -\frac{t^2}{J_K} (c_0+2c_2+c_3), \label{eq:J1-square}
\ee
while the second and third-nearest neighbor couplings are given by 
\be
\tilde J_2 =  \frac{t^2}{J_K} c_2 , \qquad   \tilde J_3 =  \frac{t^2}{2J_K}c_3. \label{eq:J2-J3-square}
\ee
These expressions make clear that the effective couplings beyond first nearest neighbors are a result of $t/J_K$ corrections. For the triangular lattice we find that $\tilde J_1$ is given by
\be
 \tilde J^\triangle_1 =  \frac12 c_1 t -\frac{t^2}{J_K} (c_0+c_1+2c_2+c_3), \label{eq:J1-tri}
\ee
while the further-neighbor couplings take the same form as in Eq.~\eqref{eq:J2-J3-square}, with the important stipulation that the appropriate triangular lattice expressions for $c_2$ and $c_3$ should be used.

The effective exchange couplings are shown as a function of filling fraction $x$ in Fig.~\ref{fig:J_i-2D}, for both the square lattice (panel a) and triangular lattice (panel b). The general behavior is similar to what was found for the 1D toy model. Note in particular that $\tilde J_1$ changes sign as a function of $x$, reflecting the aforementioned shift towards favoring antiferromagnetism as $x\rightarrow 1$.

We can also determine the effective spin stiffness $\kappa$, which is defined by expanding the spin wave dispersion in small momenta as
\be
S\omega_\bq \approx \kappa q^2 \label{eq:kappa}.
\ee
The spin stiffness has contributions from all effective couplings $\tilde J_{1,2,3}$, and for the square lattice we find 
\be
 \kappa_{\Box} = \frac12 c_1t - \frac{t^2}{J_K} (c_0-c_3).
\ee
For the triangular lattice we obtain the result
\be
 \kappa_{\triangle} = \frac34 c_1t - \frac{3t^2}{2J_K} (c_0+c_1 - c_2-c_3).
\ee
The dimensionless integrals $c_i$ are given in Appendix~\ref{app:integrals}, where we also provide expansions of the integrals in terms of the filling $x$, which yield expressions for the stiffness as a function of electron density. For the square lattice we find 
\be\label{eq:kappa-square}
\kappa_{\Box} \simeq \frac{x}{2}\left(1 - \frac{\pi x}{2} + \frac{\pi^2 x^2}{12}\right)t
\ee
while for the triangular lattice we find
\be\label{eq:kappa-triangle}
\kappa_{\triangle} \simeq \frac{3t}{4}x\left(1 - \frac{\pi x}{\sqrt{3}} + \frac{\pi^2 x^2}{9}\right) +  \frac{t^2}{J_K}\pi  x^2 \left(4 \pi  x-3 \sqrt{3}\right)
\ee

\section{Ferromagnets and spin-orbit coupling \label{sec:FM-SOC}}

We now turn to Kondo lattice ferromagnets in the presence of spin-orbit coupling. Spin-orbit coupling is naturally present in any material and ties the orbital motion of electrons to their spin. When spin-orbit coupling is included in the description, spatial and spin degrees of freedom transform jointly under the symmetries of the crystal lattice, and the full spin rotation symmetry present in the idealized limit of vanishing spin-orbit coupling is removed. In the context of magnetism this is well-known to give rise to magnetocrystalline anisotropies, which are reflected in the spin-wave spectrum, for instance as a magnon gap. Here we examine such magnetocrystalline anisotropy effects in the case of itinerant Kondo lattice magnets. In this section, we first consider a 1D toy model and in the next section we then consider the honeycomb lattice in 2D.

\begin{figure}
	\includegraphics[scale = 1]{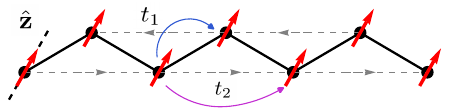}
	\caption{Model of the itinerant zigzag chain ferromagnet with spin-orbit coupling. The zigzag structure gives rise to two sublattices (labeled $A$ and $B$ in the text). The model includes a nearest and next-nearest neighbor hopping $t_1$ and $t_2$, respectively. The gray dashed lines indicate a next-nearest neighbor spin-orbit coupling $\blambda_{ij}  = \lambda \nu_{ij} \hat\bz$, with $\blambda_{ij}  $ defined in Eq.~\eqref{eq:SOC-model}. Here $\nu_{ij} = -\nu_{ji} = \pm 1$ denotes the sign structure of the spin-dependent hopping, which is opposite on the two sublattices (see text) and is indicated in the figure by arrows. Spin-orbit coupling introduces an easy-axis Ising anisotropy, such that the spins (shown as red arrows) point along the $\hat \bz$ axis. }
	\label{fig:1D-SOC}
\end{figure}

In general, when spin-orbit is included, the electronic hopping Hamiltonian takes the more general form
\be
H_{t,\lambda} = \sum_{ij} t_{ij}c^\dagger_i c_j + i \sum_{ ij } \blambda_{ij} \cdot c^\dagger_{i } \bsigma c_{j}, \label{eq:SOC-model}
\ee
where the first term describes the spin-rotation invariant piece already included in Eq.~\eqref{eq:H_klm}, and the second term describes the coupling of orbital motion and spin, as captured by the spin-dependent hoppings $\blambda_{ij}$. These hopping amplitudes satisfy $t_{ij} = t_{ji}$ and $\blambda_{ij}=-\blambda_{ji}$, which is required by Hermiticity. In the preceding companion paper \cite{strockoz2024} we have detailed how the effect of spin-orbit coupling can be incorporated in the strong coupling wave expansion in a straightforward way. With spin-orbit coupling included, the Hamiltonian from which the spin wave spectrum can be calculated in the strong coupling limit is given by
\be
\mathcal H_0 = \frac{1}{4S} \sum_{ij} (2 t^{\down\down}_{ij}a^\dagger_ia_j -t^{\up\up}_{ij} a^\dagger_ia_i-t^{\up\up}_{ij} a^\dagger_ja_j)f^\dagger_{i}f_{j}. \label{eq:H_0-FM-SOC}
\ee
This Hamiltonian is the generalization of Eq.~\eqref{eq:H_0-FM}. The spin-projected hoppings $t^{\sigma \sigma'}_{ij}$ are given by $t^{\sigma \sigma'}_{ij} = [U^\dagger_i (t_{ij} + i \blambda_{ij} \cdot  \bsigma)U_j ]_{\sigma\sigma'}$ and capture two effects. The first is a change to a local site-dependent spin quantization axis aligned with ordered local moments. This change of basis is encoded in the unitary rotations $U_i$. The second is the spin-dependent hopping coming from spin-orbit coupling, which is written in matrix form as $t_{ij} + i \blambda_{ij} \cdot  \bsigma$. When spin-orbit coupling is absent, the ordered moments of a ferromagnet can always be chosen along the $\hat z$ axis, i.e., the natural spin quantization axis. This is no longer true in the presence of spin-orbit coupling, and one must therefore rely on Eq.~\eqref{eq:H_0-FM-SOC} to compute the spin wave dispersion in the strong coupling limit. In the two models considered in this section and the next section, we will find that the ferromagnet preferentially orders along the $\hat z$ axis due to easy-axis anisotropy, and difference between $t^{\down\down}_{ij}$ and $t^{\up\up}_{ij}$ is therefore entirely due to spin-orbit coupling.

The central result we will demonstrate in this section---and which also applies to the honeycomb lattice model in the next section---is that the spin wave dispersion of a spin-orbit coupled itinerant Kondo-lattice ferromagnet is equivalent to the spin wave dispersion of a Heisenberg ferromagnet with DM and Ising magneto-crystalline anisotropy. A Heisenberg ferromagnet of this kind is described by a spin model of the form
\begin{multline}
H = -J \sum_{\langle ij \rangle} \bS_i\cdot \bS_j + \sum_{\langle\langle ij \rangle\rangle} \bD_{ij}\cdot  \bS_i\times \bS_j  \\
- \sum_{\langle\langle ij \rangle\rangle} [K^{xx}(S^x_iS^x_j+S^y_iS^y_j)+K^{zz}S^z_iS^z_j], \label{eq:H-HB-SOC}
\end{multline}
where $J$ is a nearest neighbor (ferromagnetic) exchange coupling, $\bD_{ij}$ is a next-nearest neighbor DM interaction  ~\cite{MoriyaPhysRev120, DZYALOSHINSKYJPhysChemSolids1958}, and $K^{xx}$ and $K^{zz}$ are next-nearest neighbor exchange couplings. We will show that the leading order ferromagnetic spin wave dispersion obtained from the electronic models considered below exactly maps to the linear spin wave dispersion computed from Eq.~\eqref{eq:H-HB-SOC}, and we will provide expressions for the effective exchange couplings in terms of parameters of the electronic models.

\subsection{Zigzag chain with Ising anisotropy}
\label{ssec:zigzag}

The remainder of this section is devoted to an analysis of the one-dimensional (1D) zigzag chain model shown in Fig.~\ref{fig:1D-SOC}. This model can be viewed as an ordinary 1D chain along the $\hat x$ direction with buckling in the $\hat y$ direction. The buckling of atomic sites gives rise to two sublattices and furthermore activates an intra-sublattice spin-orbit coupling  \cite{Yanase:2014p014703,Hayami:2015p064717,Yatsushiro:2022p155157,Venderbos:arXiv2025}.

In the notation of Eq.~\eqref{eq:SOC-model} this spin-orbit coupling is a next-nearest neighbor hopping with $\blambda_{ij} $ given by $\blambda_{ij}  = \lambda \nu_{ij} \hat\bz$. Here $\nu_{r,r-\hat x}  = \pm 1$ for the $A$ ($+1$) and $B$ ($-1$) sublattice, as required by the symmetries of the buckled chain (see Fig.~\ref{fig:1D-SOC}). Hence, the form of the spin-orbit coupling term is similar to the Kane-Mele spin-orbit coupling on the honeycomb lattice (see also Sec.~\ref{sec:honeycomb}).

Including a nearest neighbor and next-nearest neighbor hopping ($t_1$ and $t_2$, respectively), as well as the spin-orbit coupling $\lambda$, the electronic Hamiltonian can be written as
\be
H^\sigma_\bk = -\begin{pmatrix} 2 t_2 \gamma'_k -2\sigma\lambda \xi'_k &  2t_1 \gamma_k \\ 2t_1 \gamma_k &2 t_2 \gamma'_k +2\sigma\lambda \xi'_k\end{pmatrix}, \label{eq:H_k-1D}
\ee
where $\sigma = \up,\down$ labels spin, and we have defined and used the form factors
\begin{gather}
\gamma_k = \cos(k/2), \quad \xi_k =\sin(k/2), \label{eq:1D-nn} \\
\gamma'_k = \cos k, \quad \xi'_k =\sin k. \label{eq:1D-nnn}
\end{gather}
Introducing these form factor functions is convenient for what follows. In keeping with the convention of previous sections, $\gamma_k$ and $\xi_k$ refer to nearest neighbor couplings, whereas their primed variants refer to next-nearest neighbor couplings. 

Next, we include the Kondo coupling to a classical ferromagnet in the strong coupling limit. The fermionic part of the Hamiltonian then takes the form
\be
\mathcal H_f =  \sum_{ij}( t^{\up\up}_{ij} -\mu \delta_{ij}) f^\dagger_{i}f_{j}, \label{eq:H_f-SOC}
\ee
where $f_i$ are effectively spinless operators corresponding to electrons with their spin tied to the direction of the local moments. As mentioned, in this spin-orbit coupled case the effective hoppings $t^{\up\up}_{ij}$ depend on the direction of the ferromagnetic local moments, and the electronic contribution to the ground state energy therefore also depends on this direction. To find the ground state around which to perform a spin wave expansion, the energy of Eq.~\eqref{eq:H_f-SOC} must be minimized with respect to the direction of the local moments. We find that the energy is minimized when the local moments are oriented along the $\hat \bz$ direction. From this result we draw the conclusion that spin-orbit coupling introduces easy-axis magnetic anisotropy. In what follows we choose the local moments in the $+ \hat\bz$ direction, such that the electronic Hamiltonian in the strong coupling limit [i.e., Eq.~\eqref{eq:H_f-SOC}] is simply given by $H^\up_k$ in Eq.~\eqref{eq:H_k-1D}.

\subsection{Linear spin wave theory}
\label{ssec:1D-SOC-LSW}

We now proceed to the main result of this section: the leading order strong coupling spin wave dispersion of the itinerant spin-orbit coupled ferromagnet. The spin wave spectrum is calculated from Eq.~\eqref{eq:H_0-FM-SOC} by first averaging over the fermions in the ground state, and then determining the spin wave dispersion from the resulting quadratic boson Hamiltonian. 

As mentioned, the bosonic spin wave Hamiltonian is equivalent to a linear spin wave expansion of Eq.~\eqref{eq:H-HB-SOC} around the classical ferromagnet. To discuss the spin wave dispersion of the itinerant magnet, we will therefore first discuss the linear spin wave expansion of Eq.~\eqref{eq:H-HB-SOC}. The spin wave dispersion of the Heisenberg ferromagnet is expressed in terms of the couplings $J$, $K^{xx}$, $K^{zz}$, and $D$, such that the dispersion of the itinerant model can simply be stated by defining corresponding effective coupling constants (i.e., $\tilde J$, $\tilde D$, $\tilde K^{xx}$, $\tilde K^{zz}$) in terms of $t_1$, $t_2$, and $\lambda$. 

\begin{figure}
    \centering
    \includegraphics[scale = 0.9]{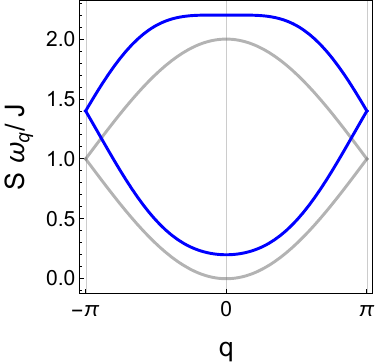}
    \caption{Plot of the magnon spectrum $\omega^\pm_q$ of the 1D zigzag chain, as given by Eq.~\eqref{eq:1d-dispersion}, for different choices of exchange couplings $J$, $K^{xx}$, $K^{zz}$, and $D$ [as defined in Eq.~\eqref{eq:H-HB-SOC}]. Gray lines correspond to $D = K^{xx} = K^{zz} = 0$, while blue lines correspond to $D = 0.2J$, $K^{zz} = 0.15J$, and $K^{xx} = 0.05J$. }
    \label{fig:magnon-spectrum-1D}
\end{figure}

After making the substitution $J \rightarrow J/S^2$ (and similarly for the other couplings) the linear spin wave Hamiltonian obtained from a $1/S$ expansion of Eq.~\eqref{eq:H-HB-SOC} around a ferromagnet along the $+ \hat\bz$ direction can be written as
\be
\mathcal H_{\text{LSW}} = \frac{1}{S} \sum_q (a^\dagger_q, b^\dagger_q)\mathcal H_q \begin{pmatrix} a_q \\ b_q\end{pmatrix}, \label{eq:H-boson-AB}
\ee
where $a_q$ and $b_q$ are bosonic operators corresponding to the $A$ and $B$ sublattice, respectively. The  Hamiltonian matrix $\mathcal H_q$ takes the form
\be\label{eq:Hq}
\mathcal H_q =\begin{pmatrix} A_q - B_q & C_q \\ C_q & A_q + B_q  \end{pmatrix}
\ee
where we have defined
\begin{gather}
A_q  = 2J +2K^{zz}-2K^{xx}\gamma'_q, \\
 B_q  = 2D \xi'_q ,\quad C_q = -2J \gamma_q.\label{eq:BCq}
\end{gather}
The $q$-dependent form factors are defined in Eqs.~\eqref{eq:1D-nn} and~\eqref{eq:1D-nnn}. Diagonalization of $\mathcal H_q$ is straightforward and one finds two magnon bands with energy
\be \label{eq:1d-dispersion}
S \omega^\pm_q = A_q \pm \sqrt{ B^2_q + C^2_q}.
\ee
In Fig.~\ref{fig:magnon-spectrum-1D} we show the spin wave dispersion for two sets of exchange couplings. The gray curves correspond to the choice $D = K^{zz} = K^{xx} = 0$, such that Eq.~\eqref{eq:H-HB-SOC} reduces to a simple nearest neighbor Heisenberg ferromagnet. As expected, a quadratic Goldstone mode is present at $q = 0$, and at $q=\pi$ the two magnon branches cross linearly. The latter is a feature also present in the electronic dispersion and is enforced by the space group symmetry of the zigzag chain~\cite{Venderbos:arXiv2025}. 

The blue curves in Fig.~\ref{fig:magnon-spectrum-1D} correspond to $K^{zz} = 0.15J$ and $ K^{xx} = 0.05J$, and a DM coupling $D = 0.2J$. The magnon spectrum is now gapped and the gap at $q=0$ is given by 
\be\label{eq:omegaqzero}
S \omega^-_{q=0} = 2K^{zz}-2K^{xx}.
\ee
As expected, the gap is directly proportional to the Ising anisotropy. Note that the magnon bands still cross linearly at $q=\pi$, since this crossing is mandated by the space group symmetries of the zigzag chain. 

We then proceed to the discussion of the effective couplings obtained for the itinerant ferromagnet. These are calculated by averaging over the fermions in the classical ground state (i.e., the classical ferromagnet of local moments) and are therefore expressed in terms of integrals over the fermion dispersion, as was the case in Sec.~\ref{sec:FM}.

Consider first the effective coupling $\tilde J$, for which we find
\be \label{eq:I_1}
\tilde J = \frac{t_1}{4} I_1, \quad I_1 = \frac{1}{N}\sum_k \frac{\gamma^2_k(n_1-n_2)}{\sqrt{\gamma^2_k+( \lambda \xi'_k/t_1)^2} }   
\ee
Here $I_1$ is a dimensionless integral over the BZ and $N$ is the total number of unit cells. Furthermore, $n_1 = n_{k,1}$ and $n_2 = n_{k,2}$ are the occupation numbers of the lower and upper band corresponding to Eq.~\eqref{eq:H_k-1D} (with $\sigma=\up$), respectively. Since the occupation numbers depend on electron density, $I_1$ is a function of electron filling. The dependence of the effective couplings on filling is exactly as in Sec.~\ref{sec:FM}. As expected from Sec.~\ref{sec:FM}, here we find that $\tilde J$ is proportional to $t_1$ and depends on $\lambda$ only indirectly via the denominator in $I_1$. 

This is different for the effective next-nearest neighbor couplings $\tilde K^{zz}$ and $\tilde K^{xx}$. We find that these are given by
\be\label{eq:Ks1D}
\tilde K^{zz} = \frac{t_2}{4} I_2 +  \frac{\lambda^2}{4t_1}I_3, \; \tilde K^{xx} = \frac{t_2}{4} I_2 - \frac{\lambda^2}{4t_1}I_3
\ee
where we have defined the dimensionless integrals $I_2 $ and $ I_3$ as
\be\label{eq:I_23}
I_2 =  \frac{1}{N}\sum_k \gamma'_k (n_1+n_2), \; I_3 = \frac{1}{N}\sum_k \frac{(\xi'_k)^2(n_1-n_2)}{\sqrt{\gamma^2_k+( \lambda \xi'_k/t_1)^2} }.
\ee
The expressions for the couplings $\tilde K^{zz}$ and $\tilde K^{xx}$ show that the next-nearest neighbor hopping $t_2$ generates a fully isotropic ferromagnetic Heisenberg coupling between next-nearest neighbor spins, as expected. The Ising anisotropy, defined as the difference $\tilde K^{zz} -\tilde K^{xx} $, is due to the spin-orbit coupling---also as expected. Furthermore, the Ising anisotropy is proportional to $\lambda^2$, i.e., the square of the spin-orbit coupling strength, and thus insensitive to the sign of $\lambda$. This is in agreement with the intuition that symmetric exchange anisotropies quadratically depend on spin-orbit coupling. Note also that  $\tilde K^{zz} -\tilde K^{xx} $ is manifestly positive, consistent with easy-axis anisotropy and thus with Eq.~\eqref{eq:omegaqzero}.

Finally, we find for the effective DM interaction $\tilde D$:
\be\label{eq:Deff1D}
\tilde D = \frac{\lambda}{4}(  I_2 + t_2 I_3/t_1 ).
\ee
The DM interaction is linear in spin-orbit coupling, as expected for antisymmetric exchange. We may therefore summarize the effective exchange couplings as follows. The hoppings $t_1$ and $t_2$ simply generate effective isotropic ferromagnetic nearest and next-nearest neighbor exchange couplings, which is in agreement with expectation for itinerant Kondo lattice ferromagnets (and the results of Sec.~\ref{sec:FM}). The spin-orbit coupling introduces an effective easy-axis Ising anisotropy (quadratic in $\lambda$) and an effective DM interaction (linear in $\lambda$). 
 
\begin{figure}
    \centering
    \includegraphics[scale = 0.9]{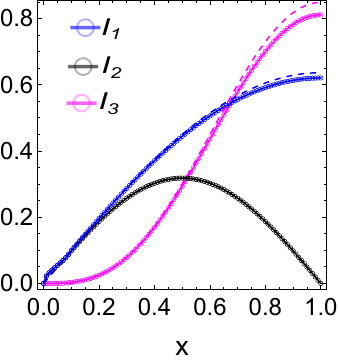}
            \caption{Numerical evaluation of the dimensionless integrals $I_{1,2,3} $ defined in Eqs.~\eqref{eq:I_1} and \eqref{eq:I_23}  as a function of filling fraction $x$ (solid lines). Here we have used $\lambda /t_1= 0.2$. Approximations to the integrals, as given by Eq.~\eqref{eq:Iapprox1D}, are shown as dashed lines. Note that in this plot $x = 1$ corresponds to a fully filled lower electron band, or one electron per unit cell (i.e., one electron per two sites). }
    \label{fig:effectivecouplings}
\end{figure}

Next, we examine the dependence of the dimensionless integrals $I_{1, 2, 3}$ on electron filling $x$, where $x$ denotes the number of electrons per unit cell (i.e., per two sites). Hence $x=1$ corresponds to the situation when the lower electron band is completely filled (i.e., $n_{k,1} = 1$ for all $k$; $n_{k,2} = 0$ for all $k$ and all shown values of $x$). We evaluate these integrals numerically and show the result in Fig.~\ref{fig:effectivecouplings} for the choice $\lambda/t_1 = 0.2$. (Note that $I_1$ and $I_3$ have a weak dependence on $\lambda$ via the electron dispersion in the denominator; $I_2$ does not depend on $\lambda$.) In the limit where $t_2 =0 $ the integrals provide direct insight into the effective couplings, since $\tilde{J} \sim I_1$, $\tilde{D} \sim I_2$, and $\tilde{K}^{zz}  = -\tilde{K}^{xx} \sim I_3$ in this case.

The integrals $I_1 $ and $I_3$ can be performed analytically when $\lambda$ is set to zero in the denominator, in which case we find
\be\label{eq:Iapprox1D} 
I_1 \simeq \frac{2}{\pi}\sin{\frac{\pi x}{2}}, \quad I_3 \simeq \frac{8}{3\pi}\sin^{3}{\frac{\pi x}{2}}. 
\ee
The integral $I_2$ can always be taken analytically and we find $I_2 =(1/\pi)\sin{\pi x} $. These approximations for $I_1$ and $I_3$ are shown by the dashed curves in Fig.~\ref{fig:effectivecouplings}, showing good agreement with the numerical results at small $\lambda/t_1 = 0.2$. 

In the limit of small electron densities, when $x \ll 1$, we find that $I_1$ and $I_2$ are both proportional to $x$, whereas $I_2 \sim x^3$. This is clearly reflected in Fig.~\ref{fig:effectivecouplings} and suggests that at small electron densities the Ising anisotropy can be ignored.

\section{Kane-Mele honeycomb lattice ferromagnet}
 \label{sec:honeycomb}

A variety of paradigmatic model systems in condensed matter physics are based on the two-dimensional honeycomb lattice. The most prominent examples include the graphene model~\cite{CastroNeto:2009p109}, the Haldane model~\cite{Haldane:1988p2015}, and the Kane-Mele model~\cite{Kane:2005p226801}. Each of these models serves to describe and elucidate fundamental properties of electronic states of matter, in particular properties which derive from the topology of the electronic wave function. 

The honeycomb lattice also provides a compelling minimal model for studying topological properties of quasiparticles other than electrons, such as magnons. One of the first models shown to realize topological magnon bands is the honeycomb lattice Heisenberg ferromagnet with a symmetry-allowed second-nearest neighbor DM interaction~\cite{Owerre:2016p386001,McClarty:2021p1, KimPRL117}. The second-nearest neighbor DM interaction is allowed for the same reason the Kane-Mele spin-orbit is symmetry-allowed in the graphene model. In the honeycomb lattice spin model the nearest neighbor ferromagnetic exchange coupling favors a ferromagnetic classical ground state (which remains stable for a nonzero but small DM interaction) while, at the level of linear spin wave theory, the DM interaction produces a term in the boson Hamiltonian akin to the Haldane term~\cite{Haldane:1988p2015}. The latter leads to a splitting of the two magnon dispersion branches at the zone corner and to a nonzero Chern number for the resulting magnon bands. 

The purpose of this section is to consider a similar scenario in an itinerant honeycomb lattice magnet. Specifically, we study a Kondo lattice model on the honeycomb lattice with added Kane-Mele spin-orbit coupling. Such a model is obtained from the isotropic model of  Eq.~\eqref{eq:H_klm} by adding a spin-orbit coupling of the general form given by Eq.~\eqref{eq:SOC-model}. In this sense, the honeycomb lattice model studied in this section can be viewed as a two-dimensional variant of the zigzag chain model examined in the previous section. The analysis of the honeycomb lattice model will therefore follow a similar structure. We will show that the linear spin wave Hamiltonian obtained for the Kondo lattice Kane-Mele ferromagnet is equivalent to spin wave Hamiltonian obtained for a honeycomb lattice Heisenberg ferromagnet described by the spin model of Eq.~\eqref{eq:H-HB-SOC}. As a result, both have the same spin wave spectrum. More importantly, this also has the immediate implication that magnon bands of the Kondo lattice Kane-Mele ferromagnet carry a Chern number and are topological.

\begin{figure}
	\includegraphics[scale = 0.9]{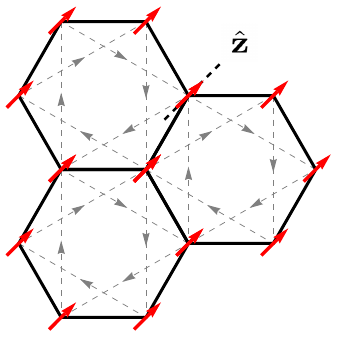}
	\caption{Honeycomb lattice Kondo model with second-nearest neighbor Kane-Mele (KM) spin-orbit coupling. Red arrows represent the ferromagnetic configuration of magnetic moments pointing in the $\hat\bz$ direction due to easy-axis anisotropy. Dashed gray lines indicate the (spin-dependent) KM hopping. The arrows indicate the direction in which $\nu_{ij} = +1$ [see \eqref{eq:KM-model} and Ref.~\onlinecite{Kane:2005p226801}]: when the arrow points from $i$ to $j$, then $ \nu_{ij} =-\nu_{ji} =1$.}
	\label{fig:honeycomb}
\end{figure}

\subsection{Kane-Mele Kondo lattice ferromagnet}

Let us first consider the honeycomb lattice Kane-Mele (KM) Hamiltonian prior to coupling the electrons to local moment spins. The KM Hamiltonian is given by~\cite{Kane:2005p226801}
\be
H_{\text{KM}} = \sum_{ij} t_{ij}c^\dagger_i c_j + i \lambda \sum_{\langle\langle ij \rangle\rangle } \nu_{ij}  c^\dagger_{i } \sigma^z c_{j} \label{eq:KM-model}
\ee
where the first term will be taken to include simple nearest and next-nearest neighbor hopping $t_1$ and $t_2$, and the second term is the KM spin-orbit coupling parameterized by $\lambda$. The sum of the second term is over next-nearest neighbors. Since the honeycomb lattice can be viewed as having a zigzag motif, the structure of this Hamiltonian is very similar to the 1D zigzag chain model. As in Sec.~\ref{ssec:zigzag}, here we have $ \nu_{ij} =-\nu_{ji}  = \pm 1$ for pairs of next-nearest neighbors $i$ and $j$ (see also Ref.~\onlinecite{Kane:2005p226801}). The precise sign structure of $\nu_{ij}$ required by symmetry is shown in Fig.~\ref{fig:honeycomb}. In Fig.~\ref{fig:honeycomb}, when the arrow on a dashed next-nearest neighbor bond points from $i$ to $j$, this indicates $ \nu_{ij} =1$. Importantly, the sign structure is opposite on the $A$ and $B$ sublattice.

As was the case in Eq.~\eqref{eq:H_k-1D}, the KM Hamiltonian can be considered separately in each spin sector labeled by $\sigma = \up,\down$, and can be written as
\be
H^\sigma_\bk = -\begin{pmatrix} 6 t_2 \gamma'_\bk -6\sigma\lambda \xi'_\bk &  3t_1 \gamma_\bk \\ 3t_1 \gamma^*_\bk &6 t_2 \gamma'_\bk +6\sigma\lambda \xi'_\bk \end{pmatrix}. \label{eq:H_k-KM}
\ee
Here we have introduced the honeycomb lattice form factor functions
\be
\gamma_\bk = \frac13 \sum_{j = 1}^3 e^{i \bk \cdot \bdelta_j},  \label{eq:gamma_k_honey}
\ee
with $\bdelta_j$ the three vectors connecting an $A$ sublattice site to its three nearest neighbors (on the $B$ sublattice), and 
\be
 \gamma'_\bk =  \frac{1}{3}\sum_{j = 1}^{3}\cos \bk \cdot \ba_j , \quad \xi'_\bk=   \frac{1}{3}\sum_{j = 1}^{3}\sin \bk \cdot \ba_j,   \label{eq:gamma-p_k_honey}
\ee
with $\ba_j$ the three primitive Bravais lattice vectors related by threefold rotation. As a result, $\gamma_\bk$ corresponds to nearest neighbor hopping, whereas $\gamma'_\bk$ and $\xi'_\bk$ correspond to next-nearest neighbor hopping. This convention  follows the treatment of the 1D zigzag chain model. 

It is well-known that Eq.~\eqref{eq:H_k-KM} describes an energy spectrum with two spin-degenerate energy bands. The KM spin-orbit coupling leads to a gap at the Brillouin zone corners. 

As in Sec.~\ref{ssec:zigzag}, we then include the Kondo coupling to classical spins and consider the limit of infinite Kondo coupling. In this limit the (classical) ground state is the ferromagnet and the Hamiltonian reduces to an effectively spinless Hamiltonian for electrons aligned with local moments. The general form of this Hamiltonian is given by Eq.~\eqref{eq:H_f-SOC}, where the effective hoppings between electrons aligned with the local moments now depend on the direction of the local moments. The latter is due to spin-orbit coupling, which ties the spin direction to the lattice. As was discussed following  Eq.~\eqref{eq:H_f-SOC}, to determine the ferromagnetic ground state it is necessary to minimize the electronic energy with respect to the direction of the local spins. As for the 1D zigzag model, here we find that the lowest energy configuration corresponds to local moments oriented along the $\hat\bz$ direction. This implies an easy-axis Ising anisotropy. In what follows, we therefore perform a spin wave expansion around the easy-axis Ising ferromagnet with moments along the $\hat\bz$ direction. The Hamiltonian for the electrons is then simply given by the $\sigma=\up$ sector in \eqref{eq:H_k-KM} and corresponds to a two-band Hamiltonian equivalent to the Haldane model. In particular, at a filling of one electron per unit cell the electrons form a Chern insulator.

\subsection{Linear spin wave theory}
\label{ssec:LSWTh}

Next, we discuss the main result of this section: the linear spin wave theory of the KM Kondo lattice ferromagnet. As mentioned, the linear spin wave Hamiltonian of the latter takes the same form as the Hamiltonian obtained from a Heisenberg ferromagnet described by Eq.~\eqref{eq:H-HB-SOC}, with appropriately defined effective coupling constants $\tilde J$, $\tilde D$, $\tilde K^{xx}$, and $\tilde K^{zz}$ (now defined on the honeycomb lattice). As in Sec.~\ref{ssec:1D-SOC-LSW}, we will proceed by first discussing the general form of the linear spin wave Hamiltonian in terms of the exchange couplings $J$, $K$, and $D$, and then examine the expressions for the effective exchange couplings, as obtained from a calculation using Eq.~\eqref{eq:H_0-FM-SOC}.

The honeycomb lattice has an $A$ and a $B$ sublattice, and the linear spin wave Hamiltonian is therefore expressed in terms of boson operators $a_\bq $ and $ b_\bq$. Its general structure is still given by Eq.~\eqref{eq:H-boson-AB}, with the difference that momentum is now $\bq = (q_x,q_y)$. The boson Hamiltonian now has the familiar form
\be\label{eq:Hqhoneycomb}
\mathcal H_\bq  =\begin{pmatrix} A_\bq  - B_\bq  & C_\bq  \\ C^*_\bq & A_\bq  + B_\bq   \end{pmatrix}
\ee
where we have defined
\begin{gather}
A_\bq  = 3J +6K^{zz}-6K^{xx}\gamma'_\bq, \\
 B_\bq  = 6 D \xi'_\bq ,\quad C_\bq = -3J \gamma_\bq.
\end{gather}
The key observation is that the structure of this boson Hamiltonian is similar to the electronic Hamiltonian given by \eqref{eq:H_k-KM}. In particular, the term $B_\bq$, which describes the DM interaction with strength $D$, enters in the same way as the Haldane term in the electronic Hamiltonian. This observation was made previously and led to the realization that a honeycomb lattice ferromagnet with DM interaction gives rise to topological magnon bands, precisely by virtue of the formal analogy with the Haldane model~\cite{Haldane:1988p2015,Owerre:2016p386001,KimPRL117,McClarty:2021p1}.

\begin{figure}
    \centering
    \includegraphics[scale = 0.9]{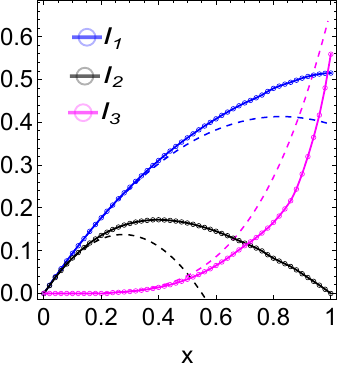}
            \caption{Numerical evaluation of the dimensionless integrals $I_{1,2,3} $ defined in Eqs.~\eqref{eq:I_1h} and \eqref{eq:I_23h} as a function of filling fraction $x$ (solid lines). Here we have used $\lambda /t_1= 0.1$ and $t_2 = 0$. Approximations to the integrals, as given by Eq.~\eqref{eq:Iapprox1D}, are shown as dashed lines. Note that in the notation used in this plot $x = 1$ corresponds to a fully filled lower band, or one electron per unit cell (i.e., per two sites).}
    \label{fig:effectivecouplings-honeycomb}
\end{figure}

The dispersion of the two magnon branches is easily obtained and given by
\be \label{eq:2d-dispersion}
S \omega^\pm_\bq = A_\bq \pm \sqrt{ B^2_\bq + |C_\bq|^2}.
\ee
In Fig.~\ref{fig:effectivecouplings-honeycomb} we show the magnon spectrum for two sets of parameter choices. The gray curves show the dispersion for a simple nearest neighbor Heisenberg ferromagnet, with $D = {K}^{zz} = {K}^{xx} = 0$. This shows the familiar and expected result: a quadratic Goldstone mode at $\bq=0$ and a linear Dirac crossing of the magnon bands at $K$ and $K'$. In contrast, when the exchange anisotropy terms are included, as shown by the blue curves, two effects are observed. The first is a magnon gap at $\bq=0$, which is proportional to ${K}^{zz} -{K}^{xx}$ and reflects the easy-axis Ising anisotropy introduced by the spin-orbit coupling. This is analogous to what we found for the zigzag chain model in Sec.~\ref{ssec:1D-SOC-LSW}. 

The second is a splitting of the magnon bands at the $K$ and $K'$ points. This splitting is proportional to the DM interaction $D$ and is given by
\be\label{eq:gapK}
\Delta_{K} \equiv S (\omega_{K}^{+} - \omega_{K}^{-}) = 2\sqrt{3}D
\ee
This splitting reflects the interpretation of $D$ as a ``Haldane gap'' for the magnons, which gives rise to two separated magnon bands with nonzero Chern number~\cite{Owerre:2016p386001,KimPRL117,McClarty:2021p1}.

\begin{figure}
    \centering
    \includegraphics[scale = 1]{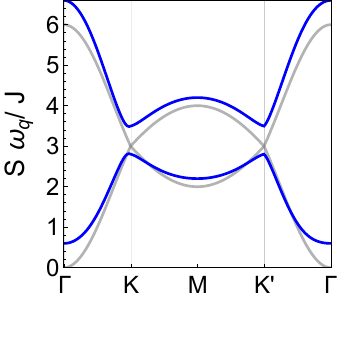}
            \caption{Magnon bands $\omega^\pm_\bq$ of the honeycomb lattice ferromagnet with easy-axis Ising anisotropy and DM interaction, as given by Eq.~\eqref{eq:2d-dispersion} and shown along the $\Gamma KMK'\Gamma$ path. The gray curves correspond to the dispersion of the simple nearest neighbor ferromagnet, with $\tilde{D} = \tilde{K}^{zz} = \tilde{K}^{xx} = 0$. The blue curves correspond to $\tilde{K}^{zz} = - \tilde{K}^{xx} = 0.05$ and $\tilde{D} = 0.2$, showing a magnon gap at $\bq=0$ and a splitting of the bands at the zone corners $K$ and $K'$.}
    \label{fig:magnonspectrum-honeycomb-spinmodel}
\end{figure}

Our next step is to determine the form of the effective exchange coupling in terms of the parameters of the Kondo lattice KM model. The spin wave Hamiltonian of the Kondo lattice KM ferromagnet is calculated from Eq.~\eqref{eq:H_0-FM-SOC} by averaging over the fermions in the classical ground state, i.e., the ferromagnetic state of classical spins along the $\hat \bz$ axis. As the ensuing discussion will show, the results are very similar to the 1D zigzag chain and therefore expose a more general structure of effective exchange couplings in itinerant spin-orbit coupled magnets.

The effective nearest neighbor exchange coupling $\tilde J$ is given by
\be\label{eq:I_1h}
\tilde J = \frac{t_1}{4} I_1, \quad I_1 = \frac{1}{N}\sum_\bk \frac{|\gamma_\bk |^2(n_1-n_2)}{\sqrt{|\gamma_\bk |^2 + (2\lambda \xi'_\bk /t_1)^2}}.
\ee
This result not only looks very similar to Eq.~\eqref{eq:I_1}, but is also consistent with the results on nearest neighbor ferromagnetic exchange of Sec.~\ref{sec:FM}. Note that $n_1 = n_{\bk,1} $ and $n_2 = n_{\bk,2}$ are the occupation factors of the lower and upper band of $H^\up_\bk$ of Eq.~\eqref{eq:H_k-KM}.

For the next-nearest neighbor effective exchange couplings $\tilde K^{zz}$ and $\tilde K^{xx}$ we find
\be\label{eq:Ktildeh}
\tilde K^{zz} = \frac{t_2}{4} I_2 +  \frac{\lambda^2}{2t_1}I_3, \quad \tilde K^{xx} = \frac{t_2}{4} I_2 - \frac{\lambda^2}{2t_1}I_3
\ee
where the dimensionless terms $I_2$ and $I_3$ are defined as
\be\label{eq:I_23h}
I_2 =  \frac{1}{N}\sum_\bk \gamma'_\bk (n_1+n_2), \; I_3 = \frac{1}{N}\sum_\bk \frac{(\xi'_\bk)^2(n_1-n_2)}{\sqrt{|\gamma_\bk |^2 + (2\lambda \xi'_\bk /t_1)^2} }.
\ee
These dimensionless integrals are straightforward 2D generalizations of Eq.~\eqref{eq:I_23}. We again find that the Ising anisotropy $\tilde K^{zz} -\tilde K^{xx} $ is proportional to $\lambda^2$ and thus quadratic in spin-orbit coupling. When $\lambda=0$ one has $\tilde{K}^{zz} = \tilde{K}^{xx}$, which simply describes isotropic next-nearest neighbor ferromagnetic exchange, whereas $t_2=0$ implies $\tilde{K}^{zz} = -\tilde{K}^{xx}$.

Lastly, we find that the effective DM coupling $\tilde{D}$ is expressed in terms of $I_2$ and $I_3$ as
\be\label{eq:Dtildeh}
\tilde D = \frac{\lambda}{4}(  I_2 + \frac{2t_2}{t_1} I_3 ),
\ee
which depends linearly on $\lambda$, as expected, and is very similar to Eq.~\eqref{eq:Deff1D}.

The dependence of the effective exchange couplings on electron density is contained in the integrals $I_{1,2,3}$. To visualize this dependence, we show $I_{1,2,3}$ as a function of filling factor $x$ in Fig.~\ref{fig:effectivecouplings-honeycomb}, where $x$ is the number of electrons per unit cell (i.e., per two sites of the honeycomb lattice). It follows from this that at low electron density the Ising anisotropy is vanishingly small and may be ignored. For small electron density the integrals may be evaluated analytically by assuming a spherical Fermi surface close to the electron band bottom. We find that for $x\ll 1$ the integrals evaluate to
\be
	I_1 \simeq x - \frac{\pi x}{3 \sqrt{3}} , \quad I_2 \simeq x - 
	\frac{\pi x^2}{\sqrt{3}}, \quad
	I_3 \simeq \frac{\pi^3 x^4}{27\sqrt{3}}
\ee
showing that both $I_1$ and $I_2$ depend linearly on $x$ to leading order, whereas $I_3 \sim x^4$. These approximations for small electron densities are indicated by dashed lines in Fig.~\ref{fig:effectivecouplings-honeycomb}, and provide a good approximation in this regime.

\mbox{}

\section{Discussion and Conclusion \label{sec:conclusion}}

In this paper we presented two applications of the canonical spin wave expansion introduced in the preceding companion paper. Both applications address general properties of itinerant Kondo lattice ferromagnets in the strong coupling regime (or double-exchange regime), and make use of key features of the canonical spin wave formalism. 

First, we have computed the first order corrections to the spin wave dispersion in the strong coupling limit. These are corrections in the strong coupling parameter $t/J_K$ and can be determined in a systematic way within the canonical spin wave expansion. We have shown that including these corrections leads to a spin wave dispersion which is equivalent to the spin wave dispersion of a Heisenberg ferromagnet with second or even third-nearest neighbor exchange interactions. At the level of linear spin wave theory this establishes a mapping between the itinerant Kondo ferromagnet and a $J_1$-$J_2$-$J_3$ Heisenberg model, even when the underlying Kondo lattice model only includes nearest neighbor electron hopping. The effective exchange couplings depend on the electron filling and can be ferromagnetic or antiferromagnetic, as has been shown for a one-dimensional toy model in one dimension and for the two-dimensional square and triangular lattices. 

Two important general conclusions follow from examining the calculated effective exchange couplings. The first is that, since the dominant nearest neighbor coupling changes sign, a transition to an antiferromagnetic state is expected as the commensurate filling of one electron per site is reached. The second is that, when a direct antiferromagnetic exchange coupling ($\jAF$) between the local moments competes with the electron-mediated exchange, the former can cause an instability towards either the antiferromagnet or an incommensurate spiral state, depending electron filling. In the strict strong coupling limit both states are part of an extensively degenerate manifold of classical states, which is reflected in an artificial full collapse of the magnon spectrum at a critical value of $\jAF$~\cite{Shannon:2002p104418,Shannon:2002p235,deGennes:1960p141}.  

In a broader sense, the mapping to an extended Heisenberg model shows that higher order corrections to the strong coupling limit introduce longer ranged interactions between the spins. This is reminiscent of RKKY physics~\cite{Ruderman:1954p99, Kasuya:1956p45, Yosida:1957p893} and is expected as one gets further away from the strong coupling regime. 

In the second part of this paper we have studied the consequences of including spin-orbit coupling effects in the description of itinerant Kondo ferromagnets in the strong coupling limit ($t/J_K \rightarrow 0$). The formulation of the canonical spin wave expansions makes the inclusion of spin-orbit coupling natural and straightforward. We have specifically examined two models of spin-orbit coupled itinerant Kondo magnets, a one-dimensional zigzag chain model and the two-dimensional honeycomb lattice model, which both feature two sublattices. Our central result, which extends beyond the two models considered, is that the linear spin wave Hamiltonian of the itinerant Kondo ferromagnet can be mapped to the spin wave Hamiltonian of a Heisenberg ferromagnetic with easy-axis Ising anisotropy and antisymmetric DM exchange interaction. 

In the case of the honeycomb lattice this mapping has an important implication: the presence of the DM interaction gives rise to topological magnon bands characterized by a nonzero Chern number. This was first pointed out in the context of spin models with magnetic anisotropy~\cite{Owerre:2016p386001,KimPRL117,Owerre:2016p043903,McClarty:2021p1} and he we have demonstrated, by virtue of the mapping to such spin models, that itinerant Kondo ferromagnets similarly feature topological magnons. A notable difference is that the itinerant ferromagnets are generally not insulating. In the model considered in this paper, the electronic sector is insulating only at commensurate filling $x=1$ (i.e., one electron per unit cell). As demonstrated, the effective anisotropic exchange couplings are electron-mediated exchange couplings and therefore sensitively depend on the electron density.

It is worth noting that spin-orbit coupling honeycomb lattice Kondo ferromagnet is not special in this regard, but is a paradigmatic example of a class of itinerant ferromagnet with topological magnon bands. Another example would be a kagome lattice ferromagnet, in full analogy with the kagome lattice Heisenberg ferromagnetic in the presence of DM interaction~\cite{Zhang:2013p144101,Mook:2014p134409,Mook:2014p024412,Chisnell:2015p147201,Owerre:2017p014422,Seshadri:2018p134411}. The itinerant kagome lattice ferromagnet with Kane-Mele spin-orbit coupling would similarly produce an effective DM interaction and thus give rise to topological magnon bands. An interesting material example of an itinerant kagome lattice ferromagnet is Fe$_3$Sn$_2$~\cite{Ye:2018p638}. We the exploration of generalizations and specific material connections for further study.

\mbox{}

\section*{Acknowledgements}

This research was supported by the National Science Foundation Award No. DMR-2144352.

\appendix

\section{Form factors and integrals \label{app:integrals}}
In this appendix, we present expressions for the form factors, and elaborate on the approximation schemes used when obtaining the effective spin stiffness for the square and triangular lattices in Section \ref{ssec:2D}.
\subsection{Square lattice}
We start by defining the form factors on the square lattice. The first, second, and third square lattice form factors are denoted by $\gamma_{\bk}, \gamma_{\bk}', \gamma_{\bk}''$, respectively and defined as below
\begin{align}
\gamma_\bk & = (\cos k_x + \cos k_y)/2 \\
\gamma'_\bk & = \cos k_x \cos k_y \\
\gamma''_\bk & = (\cos 2k_x + \cos 2k_y)/2.
\end{align}
The form factor integrals for $\gamma_\bk, \gamma_\bk', \gamma_\bk''$ are denoted by $c_1, c_2, c_3$, respectively. For brevity, we only list the form factor integral $c_1$ in Eq.~\eqref{eq:c1} and note that the two remaining integrals are obtained by substituting the corresponding form factors.
\be\label{eq:c1}
c_1 = \int_{\text{BZ}} \frac{d^2\bk}{(2\pi)^2} \gamma_\bk n_\bk
\ee
The effective spin stiffness expression $\kappa_{\Box}$ presented in Eq. ~\eqref{eq:kappa-square} is obtained by writing the relevant form factor integrals as expansions in the electron filling $x$. In this section we consider the form factor integral $c_1$, and derive the relevant expression used in the definition of the effective spin stiffness as an expansion in $x$. For small $x$, the Fermi surface can be approximated as a circle of radius $k_F$, where $k_F$ is the Fermi momentum. Under this assumption we relate the Fermi momentum to the electron filling $k_F = \sqrt{4\pi x}$.  We write the form factor integral in polar coordinates by parameterizing $k_x = k\cos{\theta}, k_y = k\sin{\theta}$, and expand for small $k$
\begin{multline}
c_1 \simeq \frac{1}{8\pi^2}\int_{0}^{k_F}\int_{-\pi}^{\pi}dk d\theta k \left(2 - \frac{k^2}{2} + \hdots\right) =  x -\frac{\pi x^2}{2}.
\end{multline}
Under the same assumptions, we find approximations of the two remaining form factor integrals
\be
c_2  \simeq x-\pi x^2, \quad c_3  \simeq x-2 \pi x^2.
\ee
\subsection{Triangular lattice}
We proceed by defining the form factors for the triangular lattice
\be
\gamma_\bk  = \frac{1}{3}\sum_{j = 1}^{3}\cos{k_j}, \; \gamma_\bk'  = \frac{1}{3}\sum_{j = 1}^{6} \cos{k_j'}, \; \gamma_\bk''  = \frac{1}{3}\sum_{j = 1}^{6} \cos{k_j''}.
\ee
Here we have defined $k_j = \bk\cdot \bm{\delta}_j$, $k'_j = \bk\cdot \bm{\delta}'_j$, and $k''_j = \bk\cdot \bm{\delta}''_j$, where $\bm{\delta}_j, \bm{\delta}_j', \bm{\delta}_j''$ are the first, second, and third-nearest neighbors of the triangular lattice. Following a similar procedure to what was described for the square lattice we approximate the form factor integrals as expansions in the electron filling $x$
\be
c_1 \simeq x - \frac{\pi x^2}{\sqrt{3}} + \frac{\pi^2 x^3}{9}
\ee
\be
c_2 \simeq x - \sqrt{3}\pi x^2 + \pi^2 x^3
\ee
\be
c_3 \simeq x - \frac{4\pi x^2}{\sqrt{3}} + \frac{16\pi^2x^3}{9}.
\ee
These are the expressions used to approximate the spin stiffness $\kappa_{\triangle}$ in Eq.~\eqref{eq:kappa-triangle}.

\mbox{}


\begin{thebibliography}{71}%
\makeatletter
\providecommand \@ifxundefined [1]{%
 \@ifx{#1\undefined}
}%
\providecommand \@ifnum [1]{%
 \ifnum #1\expandafter \@firstoftwo
 \else \expandafter \@secondoftwo
 \fi
}%
\providecommand \@ifx [1]{%
 \ifx #1\expandafter \@firstoftwo
 \else \expandafter \@secondoftwo
 \fi
}%
\providecommand \natexlab [1]{#1}%
\providecommand \enquote  [1]{``#1''}%
\providecommand \bibnamefont  [1]{#1}%
\providecommand \bibfnamefont [1]{#1}%
\providecommand \citenamefont [1]{#1}%
\providecommand \href@noop [0]{\@secondoftwo}%
\providecommand \href [0]{\begingroup \@sanitize@url \@href}%
\providecommand \@href[1]{\@@startlink{#1}\@@href}%
\providecommand \@@href[1]{\endgroup#1\@@endlink}%
\providecommand \@sanitize@url [0]{\catcode `\\12\catcode `\$12\catcode
  `\&12\catcode `\#12\catcode `\^12\catcode `\_12\catcode `\%12\relax}%
\providecommand \@@startlink[1]{}%
\providecommand \@@endlink[0]{}%
\providecommand \url  [0]{\begingroup\@sanitize@url \@url }%
\providecommand \@url [1]{\endgroup\@href {#1}{\urlprefix }}%
\providecommand \urlprefix  [0]{URL }%
\providecommand \Eprint [0]{\href }%
\providecommand \doibase [0]{https://doi.org/}%
\providecommand \selectlanguage [0]{\@gobble}%
\providecommand \bibinfo  [0]{\@secondoftwo}%
\providecommand \bibfield  [0]{\@secondoftwo}%
\providecommand \translation [1]{[#1]}%
\providecommand \BibitemOpen [0]{}%
\providecommand \bibitemStop [0]{}%
\providecommand \bibitemNoStop [0]{.\EOS\space}%
\providecommand \EOS [0]{\spacefactor3000\relax}%
\providecommand \BibitemShut  [1]{\csname bibitem#1\endcsname}%
\let\auto@bib@innerbib\@empty
\bibitem [{\citenamefont {Lovesey}(1984)}]{Lovesey1984}%
  \BibitemOpen
  \bibfield  {author} {\bibinfo {author} {\bibfnamefont {S.~W.}\ \bibnamefont
  {Lovesey}},\ }\href@noop {} {\emph {\bibinfo {title} {Theory of Neutron
  Scattering from Condensed Matter}}}\ (\bibinfo  {publisher} {Clarendon Press
  (Oxford University Press)},\ \bibinfo {address} {Oxford, UK},\ \bibinfo
  {year} {1984})\BibitemShut {NoStop}%
\bibitem [{\citenamefont {Shirane}\ \emph {et~al.}(2002)\citenamefont
  {Shirane}, \citenamefont {Shapiro},\ and\ \citenamefont
  {Tranquada}}]{Shirane2002}%
  \BibitemOpen
  \bibfield  {author} {\bibinfo {author} {\bibfnamefont {G.}~\bibnamefont
  {Shirane}}, \bibinfo {author} {\bibfnamefont {S.~M.}\ \bibnamefont
  {Shapiro}},\ and\ \bibinfo {author} {\bibfnamefont {J.~M.}\ \bibnamefont
  {Tranquada}},\ }\href@noop {} {\emph {\bibinfo {title} {Neutron Scattering
  with a Triple-Axis Spectrometer}}}\ (\bibinfo  {publisher} {Cambridge
  University Press},\ \bibinfo {address} {Cambridge, UK},\ \bibinfo {year}
  {2002})\BibitemShut {NoStop}%
\bibitem [{\citenamefont {Fishman}\ \emph {et~al.}(2018)\citenamefont
  {Fishman}, \citenamefont {Fernandez-Baca},\ and\ \citenamefont
  {R{\"o}{\"\i}m}}]{Fishman2018}%
  \BibitemOpen
  \bibfield  {author} {\bibinfo {author} {\bibfnamefont {R.~S.}\ \bibnamefont
  {Fishman}}, \bibinfo {author} {\bibfnamefont {J.~A.}\ \bibnamefont
  {Fernandez-Baca}},\ and\ \bibinfo {author} {\bibfnamefont {T.}~\bibnamefont
  {R{\"o}{\"\i}m}},\ }\href {https://doi.org/10.1088/978-1-64327-114-9} {\emph
  {\bibinfo {title} {Spin-Wave Theory and Its Applications to Neutron
  Scattering and THz Spectroscopy}}}\ (\bibinfo  {publisher} {Morgan \&
  Claypool Publishers / IOP Publishing Ltd.},\ \bibinfo {address} {Bristol,
  UK},\ \bibinfo {year} {2018})\BibitemShut {NoStop}%
\bibitem [{\citenamefont {Holstein}\ and\ \citenamefont
  {Primakoff}(1940)}]{Holstein:1940p1098}%
  \BibitemOpen
  \bibfield  {author} {\bibinfo {author} {\bibfnamefont {T.}~\bibnamefont
  {Holstein}}\ and\ \bibinfo {author} {\bibfnamefont {H.}~\bibnamefont
  {Primakoff}},\ }\bibfield  {title} {\bibinfo {title} {{Field Dependence of
  the Intrinsic Domain Magnetization of a Ferromagnet}},\ }\href
  {https://doi.org/10.1103/physrev.58.1098} {\bibfield  {journal} {\bibinfo
  {journal} {Phys. Rev.}\ }\textbf {\bibinfo {volume} {58}},\ \bibinfo {pages}
  {1098} (\bibinfo {year} {1940})}\BibitemShut {NoStop}%
\bibitem [{\citenamefont {Auerbach}(1994)}]{Auerbach1994}%
  \BibitemOpen
  \bibfield  {author} {\bibinfo {author} {\bibfnamefont {A.}~\bibnamefont
  {Auerbach}},\ }\href {https://doi.org/10.1007/978-1-4612-0869-3} {\emph
  {\bibinfo {title} {Interacting Electrons and Quantum Magnetism}}},\ Graduate
  Texts in Contemporary Physics\ (\bibinfo  {publisher} {Springer-Verlag},\
  \bibinfo {address} {New York, NY, USA},\ \bibinfo {year} {1994})\BibitemShut
  {NoStop}%
\bibitem [{\citenamefont {Zhitomirsky}\ and\ \citenamefont
  {Chernyshev}(2013)}]{ZhitomirskyRevModPhys85}%
  \BibitemOpen
  \bibfield  {author} {\bibinfo {author} {\bibfnamefont {M.~E.}\ \bibnamefont
  {Zhitomirsky}}\ and\ \bibinfo {author} {\bibfnamefont {A.~L.}\ \bibnamefont
  {Chernyshev}},\ }\bibfield  {title} {\bibinfo {title} {Colloquium:
  Spontaneous magnon decays},\ }\href
  {https://doi.org/10.1103/RevModPhys.85.219} {\bibfield  {journal} {\bibinfo
  {journal} {Rev. Mod. Phys.}\ }\textbf {\bibinfo {volume} {85}},\ \bibinfo
  {pages} {219} (\bibinfo {year} {2013})}\BibitemShut {NoStop}%
\bibitem [{\citenamefont {Strockoz}\ \emph {et~al.}(2024)\citenamefont
  {Strockoz}, \citenamefont {Frakulla}, \citenamefont {Antonenko},\ and\
  \citenamefont {Venderbos}}]{strockoz2024}%
  \BibitemOpen
  \bibfield  {author} {\bibinfo {author} {\bibfnamefont {J.}~\bibnamefont
  {Strockoz}}, \bibinfo {author} {\bibfnamefont {M.}~\bibnamefont {Frakulla}},
  \bibinfo {author} {\bibfnamefont {D.}~\bibnamefont {Antonenko}},\ and\
  \bibinfo {author} {\bibfnamefont {J.~W.~F.}\ \bibnamefont {Venderbos}},\
  }\href {https://arxiv.org/abs/2408.16665} {} (\bibinfo {year} {2024}),\
  \Eprint {https://arxiv.org/abs/2408.16665} {arXiv:2408.16665
  [cond-mat.str-el]} \BibitemShut {NoStop}%
\bibitem [{\citenamefont {Schrieffer}\ and\ \citenamefont
  {Wolff}(1966)}]{Schrieffer:1966p491}%
  \BibitemOpen
  \bibfield  {author} {\bibinfo {author} {\bibfnamefont {J.~R.}\ \bibnamefont
  {Schrieffer}}\ and\ \bibinfo {author} {\bibfnamefont {P.~A.}\ \bibnamefont
  {Wolff}},\ }\bibfield  {title} {\bibinfo {title} {{Relation between the
  Anderson and Kondo Hamiltonians}},\ }\href
  {https://doi.org/10.1103/physrev.149.491} {\bibfield  {journal} {\bibinfo
  {journal} {Phys. Rev.}\ }\textbf {\bibinfo {volume} {149}},\ \bibinfo {pages}
  {491} (\bibinfo {year} {1966})}\BibitemShut {NoStop}%
\bibitem [{\citenamefont {MacDonald}\ \emph {et~al.}(1988)\citenamefont
  {MacDonald}, \citenamefont {Girvin},\ and\ \citenamefont
  {Yoshioka}}]{MacDonald:1988p9753}%
  \BibitemOpen
  \bibfield  {author} {\bibinfo {author} {\bibfnamefont {A.~H.}\ \bibnamefont
  {MacDonald}}, \bibinfo {author} {\bibfnamefont {S.~M.}\ \bibnamefont
  {Girvin}},\ and\ \bibinfo {author} {\bibfnamefont {D.}~\bibnamefont
  {Yoshioka}},\ }\bibfield  {title} {\bibinfo {title} {{tU expansion for the
  Hubbard model}},\ }\href {https://doi.org/10.1103/physrevb.37.9753}
  {\bibfield  {journal} {\bibinfo  {journal} {Phys. Rev. B}\ }\textbf {\bibinfo
  {volume} {37}},\ \bibinfo {pages} {9753} (\bibinfo {year}
  {1988})}\BibitemShut {NoStop}%
\bibitem [{\citenamefont {Bravyi}\ \emph {et~al.}(2011)\citenamefont {Bravyi},
  \citenamefont {DiVincenzo},\ and\ \citenamefont {Loss}}]{Bravyi:2011p2793}%
  \BibitemOpen
  \bibfield  {author} {\bibinfo {author} {\bibfnamefont {S.}~\bibnamefont
  {Bravyi}}, \bibinfo {author} {\bibfnamefont {D.~P.}\ \bibnamefont
  {DiVincenzo}},\ and\ \bibinfo {author} {\bibfnamefont {D.}~\bibnamefont
  {Loss}},\ }\bibfield  {title} {\bibinfo {title} {{Schrieffer–Wolff
  transformation for quantum many-body systems}},\ }\href
  {https://doi.org/10.1016/j.aop.2011.06.004} {\bibfield  {journal} {\bibinfo
  {journal} {Annals of Physics}\ }\textbf {\bibinfo {volume} {326}},\ \bibinfo
  {pages} {2793} (\bibinfo {year} {2011})}\BibitemShut {NoStop}%
\bibitem [{\citenamefont {Chernyshev}\ \emph {et~al.}(2004)\citenamefont
  {Chernyshev}, \citenamefont {Galanakis}, \citenamefont {Phillips},
  \citenamefont {Rozhkov},\ and\ \citenamefont
  {Tremblay}}]{Chernyshev:2004p235111}%
  \BibitemOpen
  \bibfield  {author} {\bibinfo {author} {\bibfnamefont {A.~L.}\ \bibnamefont
  {Chernyshev}}, \bibinfo {author} {\bibfnamefont {D.}~\bibnamefont
  {Galanakis}}, \bibinfo {author} {\bibfnamefont {P.}~\bibnamefont {Phillips}},
  \bibinfo {author} {\bibfnamefont {A.~V.}\ \bibnamefont {Rozhkov}},\ and\
  \bibinfo {author} {\bibfnamefont {A.-M.~S.}\ \bibnamefont {Tremblay}},\
  }\bibfield  {title} {\bibinfo {title} {{Higher order corrections to effective
  low-energy theories for strongly correlated electron systems}},\ }\href
  {https://doi.org/10.1103/physrevb.70.235111} {\bibfield  {journal} {\bibinfo
  {journal} {Phys. Rev. B}\ }\textbf {\bibinfo {volume} {70}},\ \bibinfo
  {pages} {235111} (\bibinfo {year} {2004})}\BibitemShut {NoStop}%
\bibitem [{\citenamefont {Kubo}\ and\ \citenamefont
  {Ohata}(1972)}]{Kubo:1972p21}%
  \BibitemOpen
  \bibfield  {author} {\bibinfo {author} {\bibfnamefont {K.}~\bibnamefont
  {Kubo}}\ and\ \bibinfo {author} {\bibfnamefont {N.}~\bibnamefont {Ohata}},\
  }\bibfield  {title} {\bibinfo {title} {{A Quantum Theory of Double Exchange.
  I}},\ }\href {https://doi.org/10.1143/jpsj.33.21} {\bibfield  {journal}
  {\bibinfo  {journal} {J. Phys. Soc. Jpn.}\ }\textbf {\bibinfo {volume}
  {33}},\ \bibinfo {pages} {21} (\bibinfo {year} {1972})}\BibitemShut {NoStop}%
\bibitem [{\citenamefont {Furukawa}(1996)}]{Furukawa:1996p1174}%
  \BibitemOpen
  \bibfield  {author} {\bibinfo {author} {\bibfnamefont {N.}~\bibnamefont
  {Furukawa}},\ }\bibfield  {title} {\bibinfo {title} {{Spin Excitation
  Spectrum of La1- x A xMnO3}},\ }\href {https://doi.org/10.1143/jpsj.65.1174}
  {\bibfield  {journal} {\bibinfo  {journal} {J. Phys. Soc. Jpn.}\ }\textbf
  {\bibinfo {volume} {65}},\ \bibinfo {pages} {1174} (\bibinfo {year}
  {1996})}\BibitemShut {NoStop}%
\bibitem [{\citenamefont {Nagaev}(1998)}]{Nagaev:1998p827}%
  \BibitemOpen
  \bibfield  {author} {\bibinfo {author} {\bibfnamefont {E.~L.}\ \bibnamefont
  {Nagaev}},\ }\bibfield  {title} {\bibinfo {title} {{Magnon spectrum at the
  non-RKKY indirect exchange in conducting ferromagnets}},\ }\href
  {https://doi.org/10.1103/physrevb.58.827} {\bibfield  {journal} {\bibinfo
  {journal} {Phys. Rev. B}\ }\textbf {\bibinfo {volume} {58}},\ \bibinfo
  {pages} {827} (\bibinfo {year} {1998})}\BibitemShut {NoStop}%
\bibitem [{\citenamefont {Golosov}(2000)}]{Golosev:2000p3974}%
  \BibitemOpen
  \bibfield  {author} {\bibinfo {author} {\bibfnamefont {D.~I.}\ \bibnamefont
  {Golosov}},\ }\bibfield  {title} {\bibinfo {title} {{Spin Wave Theory of
  Double Exchange Ferromagnets}},\ }\href
  {https://doi.org/10.1103/physrevlett.84.3974} {\bibfield  {journal} {\bibinfo
   {journal} {Phys. Rev. Lett.}\ }\textbf {\bibinfo {volume} {84}},\ \bibinfo
  {pages} {3974} (\bibinfo {year} {2000})}\BibitemShut {NoStop}%
\bibitem [{\citenamefont {Perkins}\ and\ \citenamefont
  {Plakida}(1999)}]{Perkins:1999p1182}%
  \BibitemOpen
  \bibfield  {author} {\bibinfo {author} {\bibfnamefont {N.~B.}\ \bibnamefont
  {Perkins}}\ and\ \bibinfo {author} {\bibfnamefont {N.~M.}\ \bibnamefont
  {Plakida}},\ }\bibfield  {title} {\bibinfo {title} {{Spin dynamics in the
  generalized ferromagnetic Kondo model for manganites}},\ }\href
  {https://doi.org/10.1007/bf02557242} {\bibfield  {journal} {\bibinfo
  {journal} {Theoretical and Mathematical Physics}\ }\textbf {\bibinfo {volume}
  {120}},\ \bibinfo {pages} {1182} (\bibinfo {year} {1999})}\BibitemShut
  {NoStop}%
\bibitem [{\citenamefont {Shannon}\ and\ \citenamefont
  {Chubukov}(2002{\natexlab{a}})}]{Shannon:2002p104418}%
  \BibitemOpen
  \bibfield  {author} {\bibinfo {author} {\bibfnamefont {N.}~\bibnamefont
  {Shannon}}\ and\ \bibinfo {author} {\bibfnamefont {A.~V.}\ \bibnamefont
  {Chubukov}},\ }\bibfield  {title} {\bibinfo {title} {{Spin-wave expansion,
  finite temperature corrections, and order from disorder effects in the double
  exchange model}},\ }\href {https://doi.org/10.1103/physrevb.65.104418}
  {\bibfield  {journal} {\bibinfo  {journal} {Phys. Rev. B}\ }\textbf {\bibinfo
  {volume} {65}},\ \bibinfo {pages} {104418} (\bibinfo {year}
  {2002}{\natexlab{a}})}\BibitemShut {NoStop}%
\bibitem [{\citenamefont {Moreo}\ \emph {et~al.}(2025)\citenamefont {Moreo},
  \citenamefont {Dagotto}, \citenamefont {Alvarez}, \citenamefont {Tohyama},
  \citenamefont {Mierzejewski},\ and\ \citenamefont
  {Herbrych}}]{MoreoRepProgPhys2025}%
  \BibitemOpen
  \bibfield  {author} {\bibinfo {author} {\bibfnamefont {A.}~\bibnamefont
  {Moreo}}, \bibinfo {author} {\bibfnamefont {E.}~\bibnamefont {Dagotto}},
  \bibinfo {author} {\bibfnamefont {G.}~\bibnamefont {Alvarez}}, \bibinfo
  {author} {\bibfnamefont {T.}~\bibnamefont {Tohyama}}, \bibinfo {author}
  {\bibfnamefont {M.}~\bibnamefont {Mierzejewski}},\ and\ \bibinfo {author}
  {\bibfnamefont {J.}~\bibnamefont {Herbrych}},\ }\bibfield  {title} {\bibinfo
  {title} {Magnon damping and mode softening in quantum double-exchange
  ferromagnets},\ }\href {https://doi.org/10.1088/1361-6633/add6d4} {\bibfield
  {journal} {\bibinfo  {journal} {Reports on Progress in Physics}\ }\textbf
  {\bibinfo {volume} {88}},\ \bibinfo {pages} {068001} (\bibinfo {year}
  {2025})}\BibitemShut {NoStop}%
\bibitem [{\citenamefont {Wang}(1998)}]{Wang:1998p7427}%
  \BibitemOpen
  \bibfield  {author} {\bibinfo {author} {\bibfnamefont {X.}~\bibnamefont
  {Wang}},\ }\bibfield  {title} {\bibinfo {title} {{Theory of spin waves in a
  ferromagnetic Kondo lattice model}},\ }\href
  {https://doi.org/10.1103/physrevb.57.7427} {\bibfield  {journal} {\bibinfo
  {journal} {Phys. Rev. B}\ }\textbf {\bibinfo {volume} {57}},\ \bibinfo
  {pages} {7427} (\bibinfo {year} {1998})}\BibitemShut {NoStop}%
\bibitem [{\citenamefont {Dagotto}\ \emph {et~al.}(2001)\citenamefont
  {Dagotto}, \citenamefont {Hotta},\ and\ \citenamefont
  {Moreo}}]{Dagotto:2001p1}%
  \BibitemOpen
  \bibfield  {author} {\bibinfo {author} {\bibfnamefont {E.}~\bibnamefont
  {Dagotto}}, \bibinfo {author} {\bibfnamefont {T.}~\bibnamefont {Hotta}},\
  and\ \bibinfo {author} {\bibfnamefont {A.}~\bibnamefont {Moreo}},\ }\bibfield
   {title} {\bibinfo {title} {{Colossal magnetoresistant materials: the key
  role of phase separation}},\ }\href
  {https://doi.org/10.1016/s0370-1573(00)00121-6} {\bibfield  {journal}
  {\bibinfo  {journal} {Physics Reports}\ }\textbf {\bibinfo {volume} {344}},\
  \bibinfo {pages} {1 } (\bibinfo {year} {2001})}\BibitemShut {NoStop}%
\bibitem [{Izy(2001)}]{Izyumov:2001p109}%
  \BibitemOpen
  \bibfield  {title} {\bibinfo {title} {{Double exchange model and the unique
  properties of the manganites}},\ }\href
  {https://doi.org/10.1070/pu2001v044n02abeh000840} {\bibfield  {journal}
  {\bibinfo  {journal} {Physics-Uspekhi}\ }\textbf {\bibinfo {volume} {44}},\
  \bibinfo {pages} {109} (\bibinfo {year} {2001})}\BibitemShut {NoStop}%
\bibitem [{\citenamefont {Zener}(1951)}]{Zener:1951p440}%
  \BibitemOpen
  \bibfield  {author} {\bibinfo {author} {\bibfnamefont {C.}~\bibnamefont
  {Zener}},\ }\bibfield  {title} {\bibinfo {title} {{Interaction Between the d
  Shells in the Transition Metals}},\ }\href
  {https://doi.org/10.1103/physrev.81.440} {\bibfield  {journal} {\bibinfo
  {journal} {Phys. Rev.}\ }\textbf {\bibinfo {volume} {81}},\ \bibinfo {pages}
  {440} (\bibinfo {year} {1951})}\BibitemShut {NoStop}%
\bibitem [{\citenamefont {Anderson}\ and\ \citenamefont
  {Hasegawa}(1955)}]{Anderson:1955p675}%
  \BibitemOpen
  \bibfield  {author} {\bibinfo {author} {\bibfnamefont {P.~W.}\ \bibnamefont
  {Anderson}}\ and\ \bibinfo {author} {\bibfnamefont {H.}~\bibnamefont
  {Hasegawa}},\ }\bibfield  {title} {\bibinfo {title} {{Considerations on
  Double Exchange}},\ }\href {https://doi.org/10.1103/physrev.100.675}
  {\bibfield  {journal} {\bibinfo  {journal} {Phys. Rev.}\ }\textbf {\bibinfo
  {volume} {100}},\ \bibinfo {pages} {675} (\bibinfo {year}
  {1955})}\BibitemShut {NoStop}%
\bibitem [{\citenamefont {de~Gennes}(1960)}]{deGennesPhysRev118}%
  \BibitemOpen
  \bibfield  {author} {\bibinfo {author} {\bibfnamefont {P.~G.}\ \bibnamefont
  {de~Gennes}},\ }\bibfield  {title} {\bibinfo {title} {Effects of double
  exchange in magnetic crystals},\ }\href
  {https://doi.org/10.1103/PhysRev.118.141} {\bibfield  {journal} {\bibinfo
  {journal} {Phys. Rev.}\ }\textbf {\bibinfo {volume} {118}},\ \bibinfo {pages}
  {141} (\bibinfo {year} {1960})}\BibitemShut {NoStop}%
\bibitem [{\citenamefont {Vogt}\ \emph {et~al.}(2001)\citenamefont {Vogt},
  \citenamefont {Santos},\ and\ \citenamefont {Nolting}}]{Vogt_2001}%
  \BibitemOpen
  \bibfield  {author} {\bibinfo {author} {\bibfnamefont {M.}~\bibnamefont
  {Vogt}}, \bibinfo {author} {\bibfnamefont {C.}~\bibnamefont {Santos}},\ and\
  \bibinfo {author} {\bibfnamefont {W.}~\bibnamefont {Nolting}},\ }\bibfield
  {title} {\bibinfo {title} {Magnons in the ferromagnetic kondo-lattice
  model},\ }\href
  {https://doi.org/10.1002/1521-3951(200102)223:3<679::aid-pssb679>3.0.co;2-p}
  {\bibfield  {journal} {\bibinfo  {journal} {physica status solidi (b)}\
  }\textbf {\bibinfo {volume} {223}},\ \bibinfo {pages} {679–690} (\bibinfo
  {year} {2001})}\BibitemShut {NoStop}%
\bibitem [{\citenamefont {Furukawa}(1994)}]{Furukawa1994JPSJ}%
  \BibitemOpen
  \bibfield  {author} {\bibinfo {author} {\bibfnamefont {N.}~\bibnamefont
  {Furukawa}},\ }\bibfield  {title} {\bibinfo {title} {Transport properties of
  the kondo lattice model in the limit $s=\infty$ and $d=\infty$},\ }\href@noop
  {} {\bibfield  {journal} {\bibinfo  {journal} {Journal of the Physical
  Society of Japan}\ }\textbf {\bibinfo {volume} {63}},\ \bibinfo {pages}
  {3214} (\bibinfo {year} {1994})}\BibitemShut {NoStop}%
\bibitem [{\citenamefont {Shannon}\ and\ \citenamefont
  {Chubukov}(2002{\natexlab{b}})}]{Shannon:2002p235}%
  \BibitemOpen
  \bibfield  {author} {\bibinfo {author} {\bibfnamefont {N.}~\bibnamefont
  {Shannon}}\ and\ \bibinfo {author} {\bibfnamefont {A.~V.}\ \bibnamefont
  {Chubukov}},\ }\bibfield  {title} {\bibinfo {title} {{Order from disorder in
  the double-exchange model}},\ }\href
  {https://doi.org/10.1088/0953-8984/14/12/101} {\bibfield  {journal} {\bibinfo
   {journal} {J. Phys. Condens. Matter}\ }\textbf {\bibinfo {volume} {14}},\
  \bibinfo {pages} {L235 } (\bibinfo {year} {2002}{\natexlab{b}})}\BibitemShut
  {NoStop}%
\bibitem [{\citenamefont {Hayami}\ and\ \citenamefont
  {Motome}(2018)}]{Hayami:2018p137202}%
  \BibitemOpen
  \bibfield  {author} {\bibinfo {author} {\bibfnamefont {S.}~\bibnamefont
  {Hayami}}\ and\ \bibinfo {author} {\bibfnamefont {Y.}~\bibnamefont
  {Motome}},\ }\bibfield  {title} {\bibinfo {title} {{Néel- and Bloch-Type
  Magnetic Vortices in Rashba Metals}},\ }\href
  {https://doi.org/10.1103/physrevlett.121.137202} {\bibfield  {journal}
  {\bibinfo  {journal} {Phys. Rev. Lett.}\ }\textbf {\bibinfo {volume} {121}},\
  \bibinfo {pages} {137202} (\bibinfo {year} {2018})}\BibitemShut {NoStop}%
\bibitem [{\citenamefont {Okada}\ \emph {et~al.}(2018)\citenamefont {Okada},
  \citenamefont {Kato},\ and\ \citenamefont {Motome}}]{Okada:2018p224406}%
  \BibitemOpen
  \bibfield  {author} {\bibinfo {author} {\bibfnamefont {K.~N.}\ \bibnamefont
  {Okada}}, \bibinfo {author} {\bibfnamefont {Y.}~\bibnamefont {Kato}},\ and\
  \bibinfo {author} {\bibfnamefont {Y.}~\bibnamefont {Motome}},\ }\bibfield
  {title} {\bibinfo {title} {{Multiple-Q magnetic orders in Rashba-Dresselhaus
  metals}},\ }\href {https://doi.org/10.1103/physrevb.98.224406} {\bibfield
  {journal} {\bibinfo  {journal} {Phys. Rev. B}\ }\textbf {\bibinfo {volume}
  {98}},\ \bibinfo {pages} {224406} (\bibinfo {year} {2018})}\BibitemShut
  {NoStop}%
\bibitem [{\citenamefont {Banerjee}\ \emph {et~al.}(2014)\citenamefont
  {Banerjee}, \citenamefont {Rowland}, \citenamefont {Erten},\ and\
  \citenamefont {Randeria}}]{Banerjee:2014p031045}%
  \BibitemOpen
  \bibfield  {author} {\bibinfo {author} {\bibfnamefont {S.}~\bibnamefont
  {Banerjee}}, \bibinfo {author} {\bibfnamefont {J.}~\bibnamefont {Rowland}},
  \bibinfo {author} {\bibfnamefont {O.}~\bibnamefont {Erten}},\ and\ \bibinfo
  {author} {\bibfnamefont {M.}~\bibnamefont {Randeria}},\ }\bibfield  {title}
  {\bibinfo {title} {{Enhanced Stability of Skyrmions in Two-Dimensional Chiral
  Magnets with Rashba Spin-Orbit Coupling}},\ }\href
  {https://doi.org/10.1103/physrevx.4.031045} {\bibfield  {journal} {\bibinfo
  {journal} {Phys. Rev. X}\ }\textbf {\bibinfo {volume} {4}},\ \bibinfo {pages}
  {031045} (\bibinfo {year} {2014})}\BibitemShut {NoStop}%
\bibitem [{\citenamefont {Meza}\ and\ \citenamefont
  {Riera}(2014)}]{Meza:2014p085107}%
  \BibitemOpen
  \bibfield  {author} {\bibinfo {author} {\bibfnamefont {G.~A.}\ \bibnamefont
  {Meza}}\ and\ \bibinfo {author} {\bibfnamefont {J.~A.}\ \bibnamefont
  {Riera}},\ }\bibfield  {title} {\bibinfo {title} {{Magnetic and transport
  signatures of Rashba spin-orbit coupling on the ferromagnetic Kondo lattice
  model in two dimensions}},\ }\href
  {https://doi.org/10.1103/physrevb.90.085107} {\bibfield  {journal} {\bibinfo
  {journal} {Phys. Rev. B}\ }\textbf {\bibinfo {volume} {90}},\ \bibinfo
  {pages} {085107} (\bibinfo {year} {2014})}\BibitemShut {NoStop}%
\bibitem [{\citenamefont {Zhang}\ \emph {et~al.}(2020)\citenamefont {Zhang},
  \citenamefont {Ishizuka}, \citenamefont {Zhang}, \citenamefont {Halász},\
  and\ \citenamefont {Batista}}]{Zhang:2020p024420}%
  \BibitemOpen
  \bibfield  {author} {\bibinfo {author} {\bibfnamefont {S.-S.}\ \bibnamefont
  {Zhang}}, \bibinfo {author} {\bibfnamefont {H.}~\bibnamefont {Ishizuka}},
  \bibinfo {author} {\bibfnamefont {H.}~\bibnamefont {Zhang}}, \bibinfo
  {author} {\bibfnamefont {G.~B.}\ \bibnamefont {Halász}},\ and\ \bibinfo
  {author} {\bibfnamefont {C.~D.}\ \bibnamefont {Batista}},\ }\bibfield
  {title} {\bibinfo {title} {{Real-space Berry curvature of itinerant electron
  systems with spin-orbit interaction}},\ }\href
  {https://doi.org/10.1103/physrevb.101.024420} {\bibfield  {journal} {\bibinfo
   {journal} {Phys. Rev. B}\ }\textbf {\bibinfo {volume} {101}},\ \bibinfo
  {pages} {024420} (\bibinfo {year} {2020})}\BibitemShut {NoStop}%
\bibitem [{\citenamefont {Kathyat}\ \emph {et~al.}(2020)\citenamefont
  {Kathyat}, \citenamefont {Mukherjee},\ and\ \citenamefont
  {Kumar}}]{Kathyat:2020p075106}%
  \BibitemOpen
  \bibfield  {author} {\bibinfo {author} {\bibfnamefont {D.~S.}\ \bibnamefont
  {Kathyat}}, \bibinfo {author} {\bibfnamefont {A.}~\bibnamefont {Mukherjee}},\
  and\ \bibinfo {author} {\bibfnamefont {S.}~\bibnamefont {Kumar}},\ }\bibfield
   {title} {\bibinfo {title} {{Microscopic magnetic Hamiltonian for exotic spin
  textures in metals}},\ }\href {https://doi.org/10.1103/physrevb.102.075106}
  {\bibfield  {journal} {\bibinfo  {journal} {Phys. Rev. B}\ }\textbf {\bibinfo
  {volume} {102}},\ \bibinfo {pages} {075106} (\bibinfo {year}
  {2020})}\BibitemShut {NoStop}%
\bibitem [{\citenamefont {Kathyat}\ \emph {et~al.}(2021)\citenamefont
  {Kathyat}, \citenamefont {Mukherjee},\ and\ \citenamefont
  {Kumar}}]{Kathyat:2021p035111}%
  \BibitemOpen
  \bibfield  {author} {\bibinfo {author} {\bibfnamefont {D.~S.}\ \bibnamefont
  {Kathyat}}, \bibinfo {author} {\bibfnamefont {A.}~\bibnamefont {Mukherjee}},\
  and\ \bibinfo {author} {\bibfnamefont {S.}~\bibnamefont {Kumar}},\ }\bibfield
   {title} {\bibinfo {title} {{Electronic mechanism for nanoscale skyrmions and
  topological metals}},\ }\href {https://doi.org/10.1103/physrevb.103.035111}
  {\bibfield  {journal} {\bibinfo  {journal} {Phys. Rev. B}\ }\textbf {\bibinfo
  {volume} {103}},\ \bibinfo {pages} {035111} (\bibinfo {year}
  {2021})}\BibitemShut {NoStop}%
\bibitem [{\citenamefont {Kane}\ and\ \citenamefont
  {Mele}(2005)}]{Kane:2005p226801}%
  \BibitemOpen
  \bibfield  {author} {\bibinfo {author} {\bibfnamefont {C.~L.}\ \bibnamefont
  {Kane}}\ and\ \bibinfo {author} {\bibfnamefont {E.~J.}\ \bibnamefont
  {Mele}},\ }\bibfield  {title} {\bibinfo {title} {{Quantum Spin Hall Effect in
  Graphene}},\ }\href {https://doi.org/10.1103/physrevlett.95.226801}
  {\bibfield  {journal} {\bibinfo  {journal} {Phys. Rev. Lett.}\ }\textbf
  {\bibinfo {volume} {95}},\ \bibinfo {pages} {226801} (\bibinfo {year}
  {2005})}\BibitemShut {NoStop}%
\bibitem [{\citenamefont {Malki}\ and\ \citenamefont
  {Uhrig}(2020)}]{Malki:2020p20003}%
  \BibitemOpen
  \bibfield  {author} {\bibinfo {author} {\bibfnamefont {M.}~\bibnamefont
  {Malki}}\ and\ \bibinfo {author} {\bibfnamefont {G.~S.}\ \bibnamefont
  {Uhrig}},\ }\bibfield  {title} {\bibinfo {title} {{Topological magnetic
  excitations}},\ }\href {https://doi.org/10.1209/0295-5075/132/20003}
  {\bibfield  {journal} {\bibinfo  {journal} {Europhysics Letters}\ }\textbf
  {\bibinfo {volume} {132}},\ \bibinfo {pages} {20003} (\bibinfo {year}
  {2020})}\BibitemShut {NoStop}%
\bibitem [{\citenamefont {Kondo}\ \emph {et~al.}(2020)\citenamefont {Kondo},
  \citenamefont {Akagi},\ and\ \citenamefont {Katsura}}]{Kondo:2020pptaa151}%
  \BibitemOpen
  \bibfield  {author} {\bibinfo {author} {\bibfnamefont {H.}~\bibnamefont
  {Kondo}}, \bibinfo {author} {\bibfnamefont {Y.}~\bibnamefont {Akagi}},\ and\
  \bibinfo {author} {\bibfnamefont {H.}~\bibnamefont {Katsura}},\ }\bibfield
  {title} {\bibinfo {title} {{Non-Hermiticity and topological invariants of
  magnon Bogoliubov-de Gennes systems}},\ }\href
  {https://doi.org/10.1093/ptep/ptaa151} {\bibfield  {journal} {\bibinfo
  {journal} {Progress of Theoretical and Experimental Physics}\ }\textbf
  {\bibinfo {volume} {2020}},\ \bibinfo {pages} {ptaa151} (\bibinfo {year}
  {2020})}\BibitemShut {NoStop}%
\bibitem [{\citenamefont {McClarty}(2021)}]{McClarty:2021p1}%
  \BibitemOpen
  \bibfield  {author} {\bibinfo {author} {\bibfnamefont {P.~A.}\ \bibnamefont
  {McClarty}},\ }\bibfield  {title} {\bibinfo {title} {{Topological Magnons: A
  Review}},\ }\href {https://doi.org/10.1146/annurev-conmatphys-031620-104715}
  {\bibfield  {journal} {\bibinfo  {journal} {Annu. Rev. Condens. Matter
  Phys.}\ }\textbf {\bibinfo {volume} {13}},\ \bibinfo {pages} {1} (\bibinfo
  {year} {2021})}\BibitemShut {NoStop}%
\bibitem [{\citenamefont {Li}\ \emph {et~al.}(2021)\citenamefont {Li},
  \citenamefont {Cao},\ and\ \citenamefont {Yan}}]{Li:2021p1}%
  \BibitemOpen
  \bibfield  {author} {\bibinfo {author} {\bibfnamefont {Z.-X.}\ \bibnamefont
  {Li}}, \bibinfo {author} {\bibfnamefont {Y.}~\bibnamefont {Cao}},\ and\
  \bibinfo {author} {\bibfnamefont {P.}~\bibnamefont {Yan}},\ }\bibfield
  {title} {\bibinfo {title} {{Topological insulators and semimetals in
  classical magnetic systems}},\ }\href
  {https://doi.org/10.1016/j.physrep.2021.02.003} {\bibfield  {journal}
  {\bibinfo  {journal} {Physics Reports}\ }\textbf {\bibinfo {volume} {915}},\
  \bibinfo {pages} {1} (\bibinfo {year} {2021})}\BibitemShut {NoStop}%
\bibitem [{\citenamefont {Bonbien}\ \emph {et~al.}(2022)\citenamefont
  {Bonbien}, \citenamefont {Zhuo}, \citenamefont {Salimath}, \citenamefont
  {Ly}, \citenamefont {Abbout},\ and\ \citenamefont
  {Manchon}}]{Bonbien:2022p103002}%
  \BibitemOpen
  \bibfield  {author} {\bibinfo {author} {\bibfnamefont {V.}~\bibnamefont
  {Bonbien}}, \bibinfo {author} {\bibfnamefont {F.}~\bibnamefont {Zhuo}},
  \bibinfo {author} {\bibfnamefont {A.}~\bibnamefont {Salimath}}, \bibinfo
  {author} {\bibfnamefont {O.}~\bibnamefont {Ly}}, \bibinfo {author}
  {\bibfnamefont {A.}~\bibnamefont {Abbout}},\ and\ \bibinfo {author}
  {\bibfnamefont {A.}~\bibnamefont {Manchon}},\ }\bibfield  {title} {\bibinfo
  {title} {{Topological aspects of antiferromagnets}},\ }\href
  {https://doi.org/10.1088/1361-6463/ac28fa} {\bibfield  {journal} {\bibinfo
  {journal} {J. Phys. D}\ }\textbf {\bibinfo {volume} {55}},\ \bibinfo {pages}
  {103002} (\bibinfo {year} {2022})}\BibitemShut {NoStop}%
\bibitem [{\citenamefont {Katsura}\ \emph {et~al.}(2010)\citenamefont
  {Katsura}, \citenamefont {Nagaosa},\ and\ \citenamefont
  {Lee}}]{Katsura:2010p066403}%
  \BibitemOpen
  \bibfield  {author} {\bibinfo {author} {\bibfnamefont {H.}~\bibnamefont
  {Katsura}}, \bibinfo {author} {\bibfnamefont {N.}~\bibnamefont {Nagaosa}},\
  and\ \bibinfo {author} {\bibfnamefont {P.~A.}\ \bibnamefont {Lee}},\
  }\bibfield  {title} {\bibinfo {title} {{Theory of the Thermal Hall Effect in
  Quantum Magnets}},\ }\href {https://doi.org/10.1103/physrevlett.104.066403}
  {\bibfield  {journal} {\bibinfo  {journal} {Phys. Rev. Lett.}\ }\textbf
  {\bibinfo {volume} {104}},\ \bibinfo {pages} {066403} (\bibinfo {year}
  {2010})}\BibitemShut {NoStop}%
\bibitem [{\citenamefont {Matsumoto}\ \emph {et~al.}(2014)\citenamefont
  {Matsumoto}, \citenamefont {Shindou},\ and\ \citenamefont
  {Murakami}}]{Matsumoto:2014p054420}%
  \BibitemOpen
  \bibfield  {author} {\bibinfo {author} {\bibfnamefont {R.}~\bibnamefont
  {Matsumoto}}, \bibinfo {author} {\bibfnamefont {R.}~\bibnamefont {Shindou}},\
  and\ \bibinfo {author} {\bibfnamefont {S.}~\bibnamefont {Murakami}},\
  }\bibfield  {title} {\bibinfo {title} {{Thermal Hall effect of magnons in
  magnets with dipolar interaction}},\ }\href
  {https://doi.org/10.1103/physrevb.89.054420} {\bibfield  {journal} {\bibinfo
  {journal} {Phys. Rev. B}\ }\textbf {\bibinfo {volume} {89}},\ \bibinfo
  {pages} {054420} (\bibinfo {year} {2014})}\BibitemShut {NoStop}%
\bibitem [{\citenamefont {Haldane}(1988)}]{Haldane:1988p2015}%
  \BibitemOpen
  \bibfield  {author} {\bibinfo {author} {\bibfnamefont {F.~D.~M.}\
  \bibnamefont {Haldane}},\ }\bibfield  {title} {\bibinfo {title} {{Model for a
  Quantum Hall Effect without Landau Levels: Condensed-Matter Realization of
  the "Parity Anomaly"}},\ }\href {https://doi.org/10.1103/physrevlett.61.2015}
  {\bibfield  {journal} {\bibinfo  {journal} {Phys. Rev. Lett.}\ }\textbf
  {\bibinfo {volume} {61}},\ \bibinfo {pages} {2015 } (\bibinfo {year}
  {1988})}\BibitemShut {NoStop}%
\bibitem [{\citenamefont {Owerre}(2016{\natexlab{a}})}]{Owerre:2016p386001}%
  \BibitemOpen
  \bibfield  {author} {\bibinfo {author} {\bibfnamefont {S.~A.}\ \bibnamefont
  {Owerre}},\ }\bibfield  {title} {\bibinfo {title} {{A first theoretical
  realization of honeycomb topological magnon insulator}},\ }\href
  {https://doi.org/10.1088/0953-8984/28/38/386001} {\bibfield  {journal}
  {\bibinfo  {journal} {J. Phys. Condens. Matter}\ }\textbf {\bibinfo {volume}
  {28}},\ \bibinfo {pages} {386001} (\bibinfo {year}
  {2016}{\natexlab{a}})}\BibitemShut {NoStop}%
\bibitem [{\citenamefont {Kim}\ \emph {et~al.}(2016)\citenamefont {Kim},
  \citenamefont {Ochoa}, \citenamefont {Zarzuela},\ and\ \citenamefont
  {Tserkovnyak}}]{KimPRL117}%
  \BibitemOpen
  \bibfield  {author} {\bibinfo {author} {\bibfnamefont {S.~K.}\ \bibnamefont
  {Kim}}, \bibinfo {author} {\bibfnamefont {H.}~\bibnamefont {Ochoa}}, \bibinfo
  {author} {\bibfnamefont {R.}~\bibnamefont {Zarzuela}},\ and\ \bibinfo
  {author} {\bibfnamefont {Y.}~\bibnamefont {Tserkovnyak}},\ }\bibfield
  {title} {\bibinfo {title} {Realization of the haldane-kane-mele model in a
  system of localized spins},\ }\href
  {https://doi.org/10.1103/PhysRevLett.117.227201} {\bibfield  {journal}
  {\bibinfo  {journal} {Phys. Rev. Lett.}\ }\textbf {\bibinfo {volume} {117}},\
  \bibinfo {pages} {227201} (\bibinfo {year} {2016})}\BibitemShut {NoStop}%
\bibitem [{\citenamefont {Owerre}(2016{\natexlab{b}})}]{Owerre:2016p043903}%
  \BibitemOpen
  \bibfield  {author} {\bibinfo {author} {\bibfnamefont {S.~A.}\ \bibnamefont
  {Owerre}},\ }\bibfield  {title} {\bibinfo {title} {{Topological honeycomb
  magnon Hall effect: A calculation of thermal Hall conductivity of magnetic
  spin excitations}},\ }\href {https://doi.org/10.1063/1.4959815} {\bibfield
  {journal} {\bibinfo  {journal} {J. Appl. Phys.}\ }\textbf {\bibinfo {volume}
  {120}},\ \bibinfo {pages} {043903} (\bibinfo {year}
  {2016}{\natexlab{b}})}\BibitemShut {NoStop}%
\bibitem [{\citenamefont {Tokura}(2006)}]{TokuraRepProgPhys69}%
  \BibitemOpen
  \bibfield  {author} {\bibinfo {author} {\bibfnamefont {Y.}~\bibnamefont
  {Tokura}},\ }\bibfield  {title} {\bibinfo {title} {Critical features of
  colossal magnetoresistive manganites},\ }\href
  {https://doi.org/10.1088/0034-4885/69/3/R06} {\bibfield  {journal} {\bibinfo
  {journal} {Reports on Progress in Physics}\ }\textbf {\bibinfo {volume}
  {69}},\ \bibinfo {pages} {797} (\bibinfo {year} {2006})}\BibitemShut
  {NoStop}%
\bibitem [{\citenamefont {Tsunetsugu}\ \emph {et~al.}(1997)\citenamefont
  {Tsunetsugu}, \citenamefont {Sigrist},\ and\ \citenamefont
  {Ueda}}]{Tsunetsugu:1997p809}%
  \BibitemOpen
  \bibfield  {author} {\bibinfo {author} {\bibfnamefont {H.}~\bibnamefont
  {Tsunetsugu}}, \bibinfo {author} {\bibfnamefont {M.}~\bibnamefont
  {Sigrist}},\ and\ \bibinfo {author} {\bibfnamefont {K.}~\bibnamefont
  {Ueda}},\ }\bibfield  {title} {\bibinfo {title} {{The ground-state phase
  diagram of the one-dimensional Kondo lattice model}},\ }\href
  {https://doi.org/10.1103/revmodphys.69.809} {\bibfield  {journal} {\bibinfo
  {journal} {Rev. Mod. Phys.}\ }\textbf {\bibinfo {volume} {69}},\ \bibinfo
  {pages} {809} (\bibinfo {year} {1997})}\BibitemShut {NoStop}%
\bibitem [{\citenamefont {Tsunetsugu}\ \emph {et~al.}(1993)\citenamefont
  {Tsunetsugu}, \citenamefont {Sigrist},\ and\ \citenamefont
  {Ueda}}]{Tsunetsugu:1993p8345}%
  \BibitemOpen
  \bibfield  {author} {\bibinfo {author} {\bibfnamefont {H.}~\bibnamefont
  {Tsunetsugu}}, \bibinfo {author} {\bibfnamefont {M.}~\bibnamefont
  {Sigrist}},\ and\ \bibinfo {author} {\bibfnamefont {K.}~\bibnamefont
  {Ueda}},\ }\bibfield  {title} {\bibinfo {title} {{Phase diagram of the
  one-dimensional Kondo-lattice model}},\ }\href
  {https://doi.org/10.1103/physrevb.47.8345} {\bibfield  {journal} {\bibinfo
  {journal} {Phys. Rev. B}\ }\textbf {\bibinfo {volume} {47}},\ \bibinfo
  {pages} {8345} (\bibinfo {year} {1993})}\BibitemShut {NoStop}%
\bibitem [{\citenamefont {Sigrist}\ \emph {et~al.}(1991)\citenamefont
  {Sigrist}, \citenamefont {Tsunetsuga},\ and\ \citenamefont
  {Ueda}}]{Sigrist:1991p2211}%
  \BibitemOpen
  \bibfield  {author} {\bibinfo {author} {\bibfnamefont {M.}~\bibnamefont
  {Sigrist}}, \bibinfo {author} {\bibfnamefont {H.}~\bibnamefont
  {Tsunetsuga}},\ and\ \bibinfo {author} {\bibfnamefont {K.}~\bibnamefont
  {Ueda}},\ }\bibfield  {title} {\bibinfo {title} {{Rigorous results for the
  one-electron Kondo-lattice model}},\ }\href
  {https://doi.org/10.1103/physrevlett.67.2211} {\bibfield  {journal} {\bibinfo
   {journal} {Phys. Rev. Lett.}\ }\textbf {\bibinfo {volume} {67}},\ \bibinfo
  {pages} {2211} (\bibinfo {year} {1991})}\BibitemShut {NoStop}%
\bibitem [{\citenamefont {Sigrist}\ \emph {et~al.}(1992)\citenamefont
  {Sigrist}, \citenamefont {Ueda},\ and\ \citenamefont
  {Tsunetsugu}}]{Sigrist:1992p175}%
  \BibitemOpen
  \bibfield  {author} {\bibinfo {author} {\bibfnamefont {M.}~\bibnamefont
  {Sigrist}}, \bibinfo {author} {\bibfnamefont {K.}~\bibnamefont {Ueda}},\ and\
  \bibinfo {author} {\bibfnamefont {H.}~\bibnamefont {Tsunetsugu}},\ }\bibfield
   {title} {\bibinfo {title} {{Ferromagnetism of the Kondo lattice in the
  low-carrier-concentration limit}},\ }\href
  {https://doi.org/10.1103/physrevb.46.175} {\bibfield  {journal} {\bibinfo
  {journal} {Phys. Rev. B}\ }\textbf {\bibinfo {volume} {46}},\ \bibinfo
  {pages} {175} (\bibinfo {year} {1992})}\BibitemShut {NoStop}%
\bibitem [{\citenamefont {Frakulla}\ \emph {et~al.}(2024)\citenamefont
  {Frakulla}, \citenamefont {Strockoz}, \citenamefont {Antonenko},\ and\
  \citenamefont {Venderbos}}]{frakulla2024}%
  \BibitemOpen
  \bibfield  {author} {\bibinfo {author} {\bibfnamefont {M.}~\bibnamefont
  {Frakulla}}, \bibinfo {author} {\bibfnamefont {J.}~\bibnamefont {Strockoz}},
  \bibinfo {author} {\bibfnamefont {D.~S.}\ \bibnamefont {Antonenko}},\ and\
  \bibinfo {author} {\bibfnamefont {J.~W.~F.}\ \bibnamefont {Venderbos}},\
  }\href {https://arxiv.org/abs/2408.16752} {} (\bibinfo {year} {2024}),\
  \Eprint {https://arxiv.org/abs/2408.16752} {arXiv:2408.16752
  [cond-mat.str-el]} \BibitemShut {NoStop}%
\bibitem [{\citenamefont {Ruderman}\ and\ \citenamefont
  {Kittel}(1954)}]{Ruderman:1954p99}%
  \BibitemOpen
  \bibfield  {author} {\bibinfo {author} {\bibfnamefont {M.~A.}\ \bibnamefont
  {Ruderman}}\ and\ \bibinfo {author} {\bibfnamefont {C.}~\bibnamefont
  {Kittel}},\ }\bibfield  {title} {\bibinfo {title} {{Indirect Exchange
  Coupling of Nuclear Magnetic Moments by Conduction Electrons}},\ }\href
  {https://doi.org/10.1103/physrev.96.99} {\bibfield  {journal} {\bibinfo
  {journal} {Phys. Rev.}\ }\textbf {\bibinfo {volume} {96}},\ \bibinfo {pages}
  {99} (\bibinfo {year} {1954})}\BibitemShut {NoStop}%
\bibitem [{\citenamefont {Kasuya}(1956)}]{Kasuya:1956p45}%
  \BibitemOpen
  \bibfield  {author} {\bibinfo {author} {\bibfnamefont {T.}~\bibnamefont
  {Kasuya}},\ }\bibfield  {title} {\bibinfo {title} {{A Theory of Metallic
  Ferro- and Antiferromagnetism on Zener's Model}},\ }\href
  {https://doi.org/10.1143/ptp.16.45} {\bibfield  {journal} {\bibinfo
  {journal} {Prog. Theor. Phys.}\ }\textbf {\bibinfo {volume} {16}},\ \bibinfo
  {pages} {45} (\bibinfo {year} {1956})}\BibitemShut {NoStop}%
\bibitem [{\citenamefont {Yosida}(1957)}]{Yosida:1957p893}%
  \BibitemOpen
  \bibfield  {author} {\bibinfo {author} {\bibfnamefont {K.}~\bibnamefont
  {Yosida}},\ }\bibfield  {title} {\bibinfo {title} {{Magnetic Properties of
  Cu-Mn Alloys}},\ }\href {https://doi.org/10.1103/physrev.106.893} {\bibfield
  {journal} {\bibinfo  {journal} {Phys. Rev.}\ }\textbf {\bibinfo {volume}
  {106}},\ \bibinfo {pages} {893} (\bibinfo {year} {1957})}\BibitemShut
  {NoStop}%
\bibitem [{\citenamefont {Mukherjee}\ \emph {et~al.}(2021)\citenamefont
  {Mukherjee}, \citenamefont {Kathyat},\ and\ \citenamefont
  {Kumar}}]{Mukherjee:2021p134424}%
  \BibitemOpen
  \bibfield  {author} {\bibinfo {author} {\bibfnamefont {A.}~\bibnamefont
  {Mukherjee}}, \bibinfo {author} {\bibfnamefont {D.~S.}\ \bibnamefont
  {Kathyat}},\ and\ \bibinfo {author} {\bibfnamefont {S.}~\bibnamefont
  {Kumar}},\ }\bibfield  {title} {\bibinfo {title} {{Antiferromagnetic
  skyrmions and skyrmion density wave in a Rashba-coupled Hund insulator}},\
  }\href {https://doi.org/10.1103/physrevb.103.134424} {\bibfield  {journal}
  {\bibinfo  {journal} {Phys. Rev. B}\ }\textbf {\bibinfo {volume} {103}},\
  \bibinfo {pages} {134424} (\bibinfo {year} {2021})}\BibitemShut {NoStop}%
\bibitem [{\citenamefont {Moriya}(1960)}]{MoriyaPhysRev120}%
  \BibitemOpen
  \bibfield  {author} {\bibinfo {author} {\bibfnamefont {T.}~\bibnamefont
  {Moriya}},\ }\bibfield  {title} {\bibinfo {title} {Anisotropic superexchange
  interaction and weak ferromagnetism},\ }\href
  {https://doi.org/10.1103/PhysRev.120.91} {\bibfield  {journal} {\bibinfo
  {journal} {Phys. Rev.}\ }\textbf {\bibinfo {volume} {120}},\ \bibinfo {pages}
  {91} (\bibinfo {year} {1960})}\BibitemShut {NoStop}%
\bibitem [{\citenamefont
  {Dzyaloshinsky}(1958)}]{DZYALOSHINSKYJPhysChemSolids1958}%
  \BibitemOpen
  \bibfield  {author} {\bibinfo {author} {\bibfnamefont {I.}~\bibnamefont
  {Dzyaloshinsky}},\ }\bibfield  {title} {\bibinfo {title} {A thermodynamic
  theory of “weak” ferromagnetism of antiferromagnetics},\ }\href
  {https://doi.org/https://doi.org/10.1016/0022-3697(58)90076-3} {\bibfield
  {journal} {\bibinfo  {journal} {Journal of Physics and Chemistry of Solids}\
  }\textbf {\bibinfo {volume} {4}},\ \bibinfo {pages} {241} (\bibinfo {year}
  {1958})}\BibitemShut {NoStop}%
\bibitem [{\citenamefont {Yanase}(2014)}]{Yanase:2014p014703}%
  \BibitemOpen
  \bibfield  {author} {\bibinfo {author} {\bibfnamefont {Y.}~\bibnamefont
  {Yanase}},\ }\bibfield  {title} {\bibinfo {title} {{Magneto-Electric Effect
  in Three-Dimensional Coupled Zigzag Chains}},\ }\href
  {https://doi.org/10.7566/jpsj.83.014703} {\bibfield  {journal} {\bibinfo
  {journal} {J. Phys. Soc. Jpn.}\ }\textbf {\bibinfo {volume} {83}},\ \bibinfo
  {pages} {014703} (\bibinfo {year} {2014})}\BibitemShut {NoStop}%
\bibitem [{\citenamefont {Hayami}\ \emph {et~al.}(2015)\citenamefont {Hayami},
  \citenamefont {Kusunose},\ and\ \citenamefont {Motome}}]{Hayami:2015p064717}%
  \BibitemOpen
  \bibfield  {author} {\bibinfo {author} {\bibfnamefont {S.}~\bibnamefont
  {Hayami}}, \bibinfo {author} {\bibfnamefont {H.}~\bibnamefont {Kusunose}},\
  and\ \bibinfo {author} {\bibfnamefont {Y.}~\bibnamefont {Motome}},\
  }\bibfield  {title} {\bibinfo {title} {{Spontaneous Multipole Ordering by
  Local Parity Mixing}},\ }\href {https://doi.org/10.7566/jpsj.84.064717}
  {\bibfield  {journal} {\bibinfo  {journal} {J. Phys. Soc. Jpn.}\ }\textbf
  {\bibinfo {volume} {84}},\ \bibinfo {pages} {064717} (\bibinfo {year}
  {2015})}\BibitemShut {NoStop}%
\bibitem [{\citenamefont {Yatsushiro}\ \emph {et~al.}(2022)\citenamefont
  {Yatsushiro}, \citenamefont {Oiwa}, \citenamefont {Kusunose},\ and\
  \citenamefont {Hayami}}]{Yatsushiro:2022p155157}%
  \BibitemOpen
  \bibfield  {author} {\bibinfo {author} {\bibfnamefont {M.}~\bibnamefont
  {Yatsushiro}}, \bibinfo {author} {\bibfnamefont {R.}~\bibnamefont {Oiwa}},
  \bibinfo {author} {\bibfnamefont {H.}~\bibnamefont {Kusunose}},\ and\
  \bibinfo {author} {\bibfnamefont {S.}~\bibnamefont {Hayami}},\ }\bibfield
  {title} {\bibinfo {title} {{Analysis of model-parameter dependences on the
  second-order nonlinear conductivity in PT-symmetric collinear
  antiferromagnetic metals with magnetic toroidal moment on zigzag chains}},\
  }\href {https://doi.org/10.1103/physrevb.105.155157} {\bibfield  {journal}
  {\bibinfo  {journal} {Phys. Rev. B}\ }\textbf {\bibinfo {volume} {105}},\
  \bibinfo {pages} {155157} (\bibinfo {year} {2022})}\BibitemShut {NoStop}%
\bibitem [{\citenamefont {Venderbos}\ \emph {et~al.}(2025)\citenamefont
  {Venderbos}, \citenamefont {Gentile},\ and\ \citenamefont
  {Ortix}}]{Venderbos:arXiv2025}%
  \BibitemOpen
  \bibfield  {author} {\bibinfo {author} {\bibfnamefont {J.~W.~F.}\
  \bibnamefont {Venderbos}}, \bibinfo {author} {\bibfnamefont {P.}~\bibnamefont
  {Gentile}},\ and\ \bibinfo {author} {\bibfnamefont {C.}~\bibnamefont
  {Ortix}},\ }\bibfield  {title} {\bibinfo {title} {{Topological spin
  multipolization and linear magnetoelectric coupling in two-dimensional
  antiferromagnets}},\ }\href {https://arxiv.org/abs/2512.05862} {\bibfield
  {journal} {\bibinfo  {journal} {arXiv}\ } (\bibinfo {year}
  {2025})}\BibitemShut {NoStop}%
\bibitem [{\citenamefont {Neto}\ \emph {et~al.}(2009)\citenamefont {Neto},
  \citenamefont {Guinea}, \citenamefont {Peres}, \citenamefont {Novoselov},\
  and\ \citenamefont {Geim}}]{CastroNeto:2009p109}%
  \BibitemOpen
  \bibfield  {author} {\bibinfo {author} {\bibfnamefont {A.~H.~C.}\
  \bibnamefont {Neto}}, \bibinfo {author} {\bibfnamefont {F.}~\bibnamefont
  {Guinea}}, \bibinfo {author} {\bibfnamefont {N.~M.~R.}\ \bibnamefont
  {Peres}}, \bibinfo {author} {\bibfnamefont {K.~S.}\ \bibnamefont
  {Novoselov}},\ and\ \bibinfo {author} {\bibfnamefont {A.~K.}\ \bibnamefont
  {Geim}},\ }\bibfield  {title} {\bibinfo {title} {{The electronic properties
  of graphene}},\ }\href {https://doi.org/10.1103/revmodphys.81.109} {\bibfield
   {journal} {\bibinfo  {journal} {Rev. Mod. Phys.}\ }\textbf {\bibinfo
  {volume} {81}},\ \bibinfo {pages} {109 } (\bibinfo {year}
  {2009})}\BibitemShut {NoStop}%
\bibitem [{\citenamefont {Gennes}(1960)}]{deGennes:1960p141}%
  \BibitemOpen
  \bibfield  {author} {\bibinfo {author} {\bibfnamefont {P.~G.~d.}\
  \bibnamefont {Gennes}},\ }\bibfield  {title} {\bibinfo {title} {{Effects of
  Double Exchange in Magnetic Crystals}},\ }\href
  {https://doi.org/10.1103/physrev.118.141} {\bibfield  {journal} {\bibinfo
  {journal} {Phys. Rev.}\ }\textbf {\bibinfo {volume} {118}},\ \bibinfo {pages}
  {141} (\bibinfo {year} {1960})}\BibitemShut {NoStop}%
\bibitem [{\citenamefont {Zhang}\ \emph {et~al.}(2013)\citenamefont {Zhang},
  \citenamefont {Ren}, \citenamefont {Wang},\ and\ \citenamefont
  {Li}}]{Zhang:2013p144101}%
  \BibitemOpen
  \bibfield  {author} {\bibinfo {author} {\bibfnamefont {L.}~\bibnamefont
  {Zhang}}, \bibinfo {author} {\bibfnamefont {J.}~\bibnamefont {Ren}}, \bibinfo
  {author} {\bibfnamefont {J.-S.}\ \bibnamefont {Wang}},\ and\ \bibinfo
  {author} {\bibfnamefont {B.}~\bibnamefont {Li}},\ }\bibfield  {title}
  {\bibinfo {title} {{Topological magnon insulator in insulating
  ferromagnet}},\ }\href {https://doi.org/10.1103/physrevb.87.144101}
  {\bibfield  {journal} {\bibinfo  {journal} {Phys. Rev. B}\ }\textbf {\bibinfo
  {volume} {87}},\ \bibinfo {pages} {144101} (\bibinfo {year}
  {2013})}\BibitemShut {NoStop}%
\bibitem [{\citenamefont {Mook}\ \emph
  {et~al.}(2014{\natexlab{a}})\citenamefont {Mook}, \citenamefont {Henk},\ and\
  \citenamefont {Mertig}}]{Mook:2014p134409}%
  \BibitemOpen
  \bibfield  {author} {\bibinfo {author} {\bibfnamefont {A.}~\bibnamefont
  {Mook}}, \bibinfo {author} {\bibfnamefont {J.}~\bibnamefont {Henk}},\ and\
  \bibinfo {author} {\bibfnamefont {I.}~\bibnamefont {Mertig}},\ }\bibfield
  {title} {\bibinfo {title} {{Magnon Hall effect and topology in kagome
  lattices: A theoretical investigation}},\ }\href
  {https://doi.org/10.1103/physrevb.89.134409} {\bibfield  {journal} {\bibinfo
  {journal} {Phys. Rev. B}\ }\textbf {\bibinfo {volume} {89}},\ \bibinfo
  {pages} {134409} (\bibinfo {year} {2014}{\natexlab{a}})}\BibitemShut
  {NoStop}%
\bibitem [{\citenamefont {Mook}\ \emph
  {et~al.}(2014{\natexlab{b}})\citenamefont {Mook}, \citenamefont {Henk},\ and\
  \citenamefont {Mertig}}]{Mook:2014p024412}%
  \BibitemOpen
  \bibfield  {author} {\bibinfo {author} {\bibfnamefont {A.}~\bibnamefont
  {Mook}}, \bibinfo {author} {\bibfnamefont {J.}~\bibnamefont {Henk}},\ and\
  \bibinfo {author} {\bibfnamefont {I.}~\bibnamefont {Mertig}},\ }\bibfield
  {title} {\bibinfo {title} {{Edge states in topological magnon insulators}},\
  }\href {https://doi.org/10.1103/physrevb.90.024412} {\bibfield  {journal}
  {\bibinfo  {journal} {Phys. Rev. B}\ }\textbf {\bibinfo {volume} {90}},\
  \bibinfo {pages} {024412} (\bibinfo {year} {2014}{\natexlab{b}})}\BibitemShut
  {NoStop}%
\bibitem [{\citenamefont {Chisnell}\ \emph {et~al.}(2015)\citenamefont
  {Chisnell}, \citenamefont {Helton}, \citenamefont {Freedman}, \citenamefont
  {Singh}, \citenamefont {Bewley}, \citenamefont {Nocera},\ and\ \citenamefont
  {Lee}}]{Chisnell:2015p147201}%
  \BibitemOpen
  \bibfield  {author} {\bibinfo {author} {\bibfnamefont {R.}~\bibnamefont
  {Chisnell}}, \bibinfo {author} {\bibfnamefont {J.~S.}\ \bibnamefont
  {Helton}}, \bibinfo {author} {\bibfnamefont {D.~E.}\ \bibnamefont
  {Freedman}}, \bibinfo {author} {\bibfnamefont {D.~K.}\ \bibnamefont {Singh}},
  \bibinfo {author} {\bibfnamefont {R.~I.}\ \bibnamefont {Bewley}}, \bibinfo
  {author} {\bibfnamefont {D.~G.}\ \bibnamefont {Nocera}},\ and\ \bibinfo
  {author} {\bibfnamefont {Y.~S.}\ \bibnamefont {Lee}},\ }\bibfield  {title}
  {\bibinfo {title} {{Topological Magnon Bands in a Kagome Lattice
  Ferromagnet}},\ }\href {https://doi.org/10.1103/physrevlett.115.147201}
  {\bibfield  {journal} {\bibinfo  {journal} {Phys. Rev. Lett.}\ }\textbf
  {\bibinfo {volume} {115}},\ \bibinfo {pages} {147201} (\bibinfo {year}
  {2015})}\BibitemShut {NoStop}%
\bibitem [{\citenamefont {Owerre}(2017)}]{Owerre:2017p014422}%
  \BibitemOpen
  \bibfield  {author} {\bibinfo {author} {\bibfnamefont {S.~A.}\ \bibnamefont
  {Owerre}},\ }\bibfield  {title} {\bibinfo {title} {{Topological thermal Hall
  effect in frustrated kagome antiferromagnets}},\ }\href
  {https://doi.org/10.1103/physrevb.95.014422} {\bibfield  {journal} {\bibinfo
  {journal} {Phys. Rev. B}\ }\textbf {\bibinfo {volume} {95}},\ \bibinfo
  {pages} {014422} (\bibinfo {year} {2017})}\BibitemShut {NoStop}%
\bibitem [{\citenamefont {Seshadri}\ and\ \citenamefont
  {Sen}(2018)}]{Seshadri:2018p134411}%
  \BibitemOpen
  \bibfield  {author} {\bibinfo {author} {\bibfnamefont {R.}~\bibnamefont
  {Seshadri}}\ and\ \bibinfo {author} {\bibfnamefont {D.}~\bibnamefont {Sen}},\
  }\bibfield  {title} {\bibinfo {title} {{Topological magnons in a
  kagome-lattice spin system with XXZ and Dzyaloshinskii-Moriya
  interactions}},\ }\href {https://doi.org/10.1103/physrevb.97.134411}
  {\bibfield  {journal} {\bibinfo  {journal} {Phys. Rev. B}\ }\textbf {\bibinfo
  {volume} {97}},\ \bibinfo {pages} {134411} (\bibinfo {year}
  {2018})}\BibitemShut {NoStop}%
\bibitem [{\citenamefont {Ye}\ \emph {et~al.}(2018)\citenamefont {Ye},
  \citenamefont {Kang}, \citenamefont {Liu}, \citenamefont {Cube},
  \citenamefont {Wicker}, \citenamefont {Suzuki}, \citenamefont {Jozwiak},
  \citenamefont {Bostwick}, \citenamefont {Rotenberg}, \citenamefont {Bell},
  \citenamefont {Fu}, \citenamefont {Comin},\ and\ \citenamefont
  {Checkelsky}}]{Ye:2018p638}%
  \BibitemOpen
  \bibfield  {author} {\bibinfo {author} {\bibfnamefont {L.}~\bibnamefont
  {Ye}}, \bibinfo {author} {\bibfnamefont {M.}~\bibnamefont {Kang}}, \bibinfo
  {author} {\bibfnamefont {J.}~\bibnamefont {Liu}}, \bibinfo {author}
  {\bibfnamefont {F.~v.}\ \bibnamefont {Cube}}, \bibinfo {author}
  {\bibfnamefont {C.~R.}\ \bibnamefont {Wicker}}, \bibinfo {author}
  {\bibfnamefont {T.}~\bibnamefont {Suzuki}}, \bibinfo {author} {\bibfnamefont
  {C.}~\bibnamefont {Jozwiak}}, \bibinfo {author} {\bibfnamefont
  {A.}~\bibnamefont {Bostwick}}, \bibinfo {author} {\bibfnamefont
  {E.}~\bibnamefont {Rotenberg}}, \bibinfo {author} {\bibfnamefont {D.~C.}\
  \bibnamefont {Bell}}, \bibinfo {author} {\bibfnamefont {L.}~\bibnamefont
  {Fu}}, \bibinfo {author} {\bibfnamefont {R.}~\bibnamefont {Comin}},\ and\
  \bibinfo {author} {\bibfnamefont {J.~G.}\ \bibnamefont {Checkelsky}},\
  }\bibfield  {title} {\bibinfo {title} {{Massive Dirac fermions in a
  ferromagnetic kagome metal}},\ }\href {https://doi.org/10.1038/nature25987}
  {\bibfield  {journal} {\bibinfo  {journal} {Nature}\ }\textbf {\bibinfo
  {volume} {555}},\ \bibinfo {pages} {638} (\bibinfo {year}
  {2018})}\BibitemShut {NoStop}%
\end{thebibliography}
\end{document}